\documentclass[showpacs,prd,preprint,nofootinbib,showkeys,unsortedaddress]{revtex4}
\usepackage{amssymb}

\usepackage{bm}
\usepackage{amsmath}
\usepackage{graphicx}
\usepackage{subfigure}
\usepackage{physymb}
\usepackage[colorlinks=true,linktocpage=true,linkcolor=blue,citecolor=blue]{hyperref}
\usepackage[usenames,dvipsnames]{color}

\pdfoutput=1
\newcommand{\beq}{\begin{equation}}
\newcommand{\eeq}{\end{equation}}
\newcommand{\bqa}{\begin{eqnarray}}
\newcommand{\eqa}{\end{eqnarray}}
\newcommand{\bes}{\begin{subequations}}
\newcommand{\ees}{\end{subequations}}
\newcommand{\beal}{\begin{align}}

\begin{document}

\title{Studying the validity of relativistic hydrodynamics with a new exact solution of the Boltzmann equation}
\author{Gabriel Denicol}
\affiliation{Department of Physics, McGill University, 3600 University Street, Montreal,
QC H3A 2T8, Canada}
\author{Ulrich Heinz}
\author{Mauricio Martinez}
\affiliation{Department of Physics, The Ohio State University, Columbus, OH 43210, USA}
\author{Jorge Noronha}
\affiliation{Instituto de F\'isica, Universidade de S\~ao Paulo, C.P. 66318, 05315-970
S\~ao Paulo, SP, Brazil}
\author{Michael Strickland}
\affiliation{Physics Department, Kent State University, OH 44242 United States}

\begin{abstract}
We present an exact solution to the Boltzmann equation which describes a
system undergoing boost-invariant longitudinal and azimuthally symmetric 
radial expansion for arbitrary shear viscosity to entropy density ratio. This new
solution is constructed by considering the conformal map between Minkowski 
space and the direct product of three dimensional de Sitter space with a line. 
The resulting solution respects $SO(3)_q \otimes SO(1,1) \otimes Z_2$ symmetry. 
We compare the exact kinetic solution with exact solutions of the corresponding 
macroscopic equations that were obtained from the kinetic theory in ideal and 
second-order viscous hydrodynamic approximations. The macroscopic solutions 
are obtained in de Sitter space and are subject to the same symmetries used to 
obtain the exact kinetic solution.
\end{abstract}

\date{\today}
\maketitle

\section{Introduction}
\label{sec:intro} 

One of the most important cornerstones of statistical physics is the
Boltzmann equation. This equation has been extremely useful in describing
the behavior and transport properties of a dilute gas in terms of its
intrinsic microscopic dynamics. The Boltzmann equation is a partial
differential equation with a very rich and complex mathematical structure
which makes it difficult to solve it exactly by analytical means. There are
very few exact solutions to the Boltzmann equation in the scientific
literature. As a matter of fact, even for classical systems, the problem of
the existence and uniqueness of solutions to this kinetic equation has not
been completely sorted out for any collision kernel except in some
particular cases~\cite{Gressman}. Due to these limitations, there have been
different expansion schemes put forward in the literature that allow one to
obtain approximate solutions to the Boltzmann equation. Among the most
important approaches are the Chapman-Enskog and Grad's moments methods \cite{DeGroot:1980dk}. The generalization of these methods to 
relativistic kinetic theory has been a source of debate since its 
foundations. For instance, at first order the Chapman-Enskog method 
\cite{DeGroot:1980dk,Cercignani} leads to the relativistic Navier-Stokes (NS)
equations which are acausal and unstable \cite{Hiscock:1983zz,Hiscock:1985zz}. 
To address this problem, Israel and Stewart (IS) \cite{Israel:1979wp} 
generalized Grad's original idea to the relativistic case to create a causal
second order formulation of relativistic viscous hydrodynamics.%
\footnote{%
   It is also possible to formulate higher-order relativistic viscous
   hydrodynamics using the second law of thermodynamics as a guiding principle~\cite{El:2009vj} or Chapman-Enskog like methods in the relaxation time
   approximation~\cite{Jaiswal:2013npa,Jaiswal:2013vta,Bhalerao:2013pza}.} 
However, the original IS approach presents certain inconsistencies when
obtaining the fluid dynamical equations using truncated
approximations to the distribution function. A consistent framework
has been developed recently in \cite{Denicol:2010xn,Denicol:2012cn}. Despite these advances, different approximation schemes can lead to different results for key physical quantities such as the transport
coefficients \cite{DeGroot:1980dk,Denicol:2012es}. Exact solutions
to the Boltzmann equation allow one to compare and characterize the 
effectiveness of the different approximation methods. In addition, an exact
solution has the potential to shed light on the process of momentum
isotropization in a non-equilibrated system.

There are also very few exact solutions to the hydrodynamic 
equations of motion. Recently, Gubser developed a method to construct 
exact solutions to the relativistic ideal and first order NS hydrodynamical 
approximations for a conformal fluid \cite{Gubser:2010ze,Gubser:2010ui} 
undergoing simultaneously boost-invariant longitudinal and 
azimuthally symmetric (``radial'') transverse expansion (``Gubser flow'').
The Gubser solution is based on powerful symmetry considerations: 
It is symmetric under the $SO(3)_{q}\otimes SO(1,1)\otimes Z_{2}$ 
group of transformations (``Gubser group''). In Minkowski coordinates this symmetry group is not explicitly manifest, 
and hence the strategy is to make use of the conformal map between 
Minkowski space and the curved spacetime formed by the direct product of a 3-dimensional de Sitter (dS) space with a line, $dS_{3}\otimes R$, in which the Gubser symmetry {\em is} manifest 
\cite{Gubser:2010ui}: A fluid that expands in Minkowski space with 
Gubser-symmetric flow looks static in $dS_{3}\otimes R$. The 
equations of motion for the remaining hydrodynamic fields (temperature,
energy density, etc.) are much simpler in de Sitter space than in Minkowski space and easily solved. Once the solutions are known in $dS_{3}\otimes R$, 
it is straightforward to transform them back to Minkowski space. This yields 
a 1-parameter set of velocity and temperature profiles with non-trivial radial 
and time dependence, characterized by a single overall scale parameter 
\cite{Gubser:2010ze,Gubser:2010ui}. 

Recently, Gubser's solution 
was extended to second-order conformal IS hydrodynamics \cite{Marrochio:2013wla}. Furthermore, the authors of \cite{Hatta:2014gqa,Hatta:2014gga} generalized Gubser's original approach 
by considering more general conformal maps between Minkowski space and 
other curved spaces, resulting in exact solutions including vorticity and 
associated dissipative corrections to ideal hydrodynamics.

In this work we discuss how to carry out a similar program for kinetic
theory. If a very densely populated system is invariant under a 
certain group of symmetries, it is natural to ask how this symmetry 
becomes manifest in the one-particle distribution function. For instance, if 
there is homogeneity along a certain direction, say along the $x$ axis, this 
means that the distribution function $f(x^{\mu },p_{i})=f(t,y,z;p_{i})$, {\it i.e.} 
it does not depend on $x$ and the number of independent variables 
on which the solution of the Boltzmann equation can depend has correspondingly been reduced by one. One can guess that eventually, if the 
system has enough symmetries, it must be possible to solve the 
Boltzmann equation exactly. In the context of ultrarelativistic heavy-ion collisions, this strategy was first discussed by Baym \cite{Baym:1984np} who 
solved the Boltzmann equation exactly using the relaxation time 
approximation (RTA) for the collisional kernel.\footnote{See also
    Refs. \cite{Florkowski:2013lza,Florkowski:2013lya,Florkowski:2014sfa} for 
    recent extensions to conformal and non-conformal systems.} 
In this case, longitudinal boost-invariance and invariance under translations in the transverse plane were the only ingredients 
necessary to obtain the exact solution. It describes a transversely
homogeneous system expanding along the longitudinal direction with the
boost-invariant (scaling) flow profile discovered by Bjorken \cite{Bjorken:1982qr}. Despite its beauty and simplicity, this solution has limited applications since the spacetime dynamics of any spatially finite system is affected by transverse expansion, with highly nontrivial and experimentally observable consequences. 

Gubser's important achievement was to generalize Bjorken's macroscopic 
hydrodynamic solution to systems undergoing additionally transverse expansion. 
In this paper we show how this generalization can be extended to the  
microscopic level, by solving the Boltzmann equation with RTA collision term 
exactly for systems with Gubser symmetry that undergo simultaneous 
boost-invariant longitudinal and azimuthally symmetric transverse expansion. 
Our solution can be used to describe systems with any value of the shear 
viscosity to entropy density ratio $\eta/{\mathcal{S}}$, and the result can be used to test the efficiency of various macroscopic (hydrodynamic) approximation methods. In addition, 
it can help us to understand the dynamics of the 
isotropization/thermalization process in anisotropically expanding 
systems with different longitudinal and transverse expansion rates.

Some of the ideas presented in this work were already introduced by us in a previous publication~\cite{Denicol:2014xca}. In this paper we present a more detailed derivation of our exact solution as well as an extended discussion of our findings. 
The paper is organized as follows. In Sec.~\ref{sec:Gubser} we present a
short overview of the exact Gubser solution of conformal hydrodynamics. In
Sec.~\ref{sec:Boltzeqn} we discuss the necessary aspects of the Boltzmann
equation in a curved spacetime. In Sec.~\ref{sec:Gubsersol} we present the
main result of this work, our new exact solution to the Boltzmann
equation for a conformal system with Gubser symmetry. In this section we also describe how to recover first- and second-order conformal
hydrodynamics from the exact solution of the Boltzmann equation. In Sec.~\ref{sec:result} we discuss some aspects of the exact
solution, illustrate it graphically, and make comparisons with the predictions
of different approximation methods for solving the Boltzmann equation. Our conclusions are summarized in Sec.~\ref{sec:concl}.

Before proceeding to the body of this work, let us define our metric 
conventions and notations. The metric signature is taken to be ``mostly plus'': in Minkowski space the spacetime distance
between two events is written in Cartesian coordinates $x^{\mu }=(t,\bm{x})$
as
\begin{equation}
ds^{2}=g_{\mu \nu }dx^{\mu }dx^{\nu }=-dt^{2}+dx^{2}+dy^{2}+dz^{2}\,.
\end{equation}%
With this signature convention the flow velocity $u^{\mu }$ is normalized as 
$u_{\mu }u^{\mu }=-1$. Milne coordinates in Minkowski space are defined 
by $x^{\mu }=(\tau ,x,y,\varsigma )$, with longitudinal proper time
$\tau =\sqrt{t^{2}{-}z^{2}}$, spacetime rapidity $\varsigma =\atanh{(z/t)}$, 
and metric $ds^2=-d\tau^2{}+dx^2{+}dy^2{+}\tau^2d\varsigma^2$. Polar 
coordinates in the transverse plane are defined as usual by 
$r=\sqrt{x^{2}{+}y^{2}}$ and $\phi =\atan{(y/x)}$. We denote the scalar 
product between two four-vectors with a dot, {\it i.e.} 
$A_{\mu }B^{\mu }\equiv A\cdot B$.

\section{The Gubser solution of conformal hydrodynamics}
\label{sec:Gubser} 

In this section we briefly review the techniques introduced by Gubser~\cite%
{Gubser:2010ze} to find exact solutions for conformally invariant
relativistic fluid dynamics. For a more complete discussion we refer the
reader to the original works~\cite{Gubser:2010ze,Gubser:2010ui}. The
material discussed in this section provides the necessary background for 
our analogous treatment of the Boltzmann equation in Sec.~\ref{sec:Boltzeqn}.

Gubser's exact solution to the conformal hydrodynamic equations 
respects $SO(3)_q\otimes SO(1,1)\otimes Z_2$ symmetry. The $SO(3)_q$ 
group reflects invariance under rotations in the transverse plane 
coupled with two special conformal transformations. In 
the Minkowski coordinates $x^\mu=(\tau,r,\phi,\varsigma)$ the 
generators of $SO(3)_q$ are given by \cite{Gubser:2010ze} 
\begin{subequations}
\label{eq:genMinkSO3}
\begin{align}
\xi_{1}&= 2q^2 \tau r \cos \phi \frac{\partial}{\partial(q\tau)}
+\left(1{+}q^2\tau^2{+}q^2 r^2\right) \cos\phi \frac{\partial}
{\partial(q r)}-\frac{1{+}q^2
\tau^2{-}q^2 r^2}{qr}\sin\phi \frac{\partial}{\partial \phi} \, , \\
\xi_{2}&= 2q^2 \tau r \sin \phi \frac{\partial}{\partial(q\tau)}+\left(1{+}q^2
\tau^2{+}q^2 r^2\right) \sin\phi \frac{\partial}{\partial(qr)}+\frac{1{+}q^2
\tau^2{-}q^2 r^2}{qr}\cos\phi \frac{\partial}{\partial \phi} \, , \\
\xi_3 &= \frac{\partial}{\partial \phi}\,.
\end{align}
\end{subequations}
Here $q$ is an arbitrary energy scale; the solution is invariant under 
a change of $q$ if simultaneously its transverse radius $r$ and the
longitudinal proper time $\tau$ are rescaled by $1/q$. Invariance under 
$SO(1,1)$ translates into boost invariance along the longitudinal axis and its
generator is simply $\partial/\partial\varsigma$. The $Z_2$ invariance 
is associated with longitudinal reflection symmetry under 
$\varsigma \to -\varsigma$.

It is not straighforward to obtain the flow velocity profile from
the $SO(3)_{q}$ generators \eqref{eq:genMinkSO3} in Minkowski space.
However, the flow is naturally understood
\cite{Gubser:2010ze,Gubser:2010ui} in the curved spacetime $dS_{3}\otimes R$ 
which is related to Minkowski space via a Weyl-rescaling of the metric: 
\begin{equation}
d\hat{s}^2=\frac{ds^2}{\tau^2}
=\frac{-d\tau^2+dx^2+dy^2}{\tau^2}+d\varsigma^2
=\frac{-d\tau^2+dr^2+r^2 d\phi^2}{\tau^2}+d\varsigma^2\,.  
\label{metricdS3R}
\end{equation}
If one parametrizes the variables $(\tau ,r)$ in terms of new coordinates 
$(\rho ,\theta )$ defined by 
\begin{subequations}
\label{eq:rhotheta}
\begin{align}
\rho(\tau,r)& =-\mathrm{arcsinh}\left( \frac{1-q^2\tau^2+q^2r^2}
{2q\tau }\right) \,,  
\label{definerho} \\
\theta (\tau,r)& =\mathrm{arctan}\left(\frac{2qr}{1+q^2\tau^2-q^2r^2}\right) \,,  
\label{definetheta}
\end{align}
\end{subequations}
then the measure \eqref{metricdS3R} becomes 
\begin{equation}
d\hat{s}^{2}=-d\rho ^{2}+\cosh ^{2}\!\rho \left( d\theta ^{2}+\sin
^{2}\theta\, d\phi ^{2}\right) +d\varsigma ^{2}\,,  \label{eq:linedS3R}
\end{equation}
with metric $\hat{g}_{\mu\nu}=\mathrm{diag}(-1,\cosh^2\rho,%
\cosh^2\rho\,\sin^2\theta,1)$. In the new coordinate system 
$(\rho ,\theta ,\phi ,\varsigma )$ the $SO(3)_{q}$ conformal symmetry is
manifest since the measure \eqref{eq:linedS3R} is invariant under rotations of 
the sphere parametrized by $(\theta ,\phi )$. In these coordinates, the generators
of the $SO(3)_{q}$ group are given by 
\begin{subequations}
\label{eq:gendS3}
\begin{align}
\xi _{1}& =2\left( \cos \phi \frac{\partial }{\partial \theta }-\cot \theta
\sin \phi \frac{\partial }{\partial \phi }\right) \,, \\
\xi _{2}& =2\left( \sin \phi \frac{\partial }{\partial \theta }+\cot \theta
\cos \phi \frac{\partial }{\partial \phi }\right) \,, \\
\xi _{3}& =\frac{\partial }{\partial \phi }\,,
\end{align}%
\end{subequations}
which are precisely the well known angular momentum generators. In this
paper, all quantities in de Sitter coordinates are denoted with a hat. 

In $dS_{3}\otimes R$, it is straightforward to see that the flow velocity $%
\hat{u}_{\mu }=(-1,0,0,0)$ is completely invariant under the $SO(3)_{q}$
generators~\eqref{eq:gendS3}.%
\footnote{\label{fn3}%
   This can be compared with the Bjorken flow solution
   where $u_{\mu}{\,\equiv\,}(u_\tau,u_r,u_\phi,u_\varsigma){\,=\,}(-1,0,0,0)$
   appears as the only time-like unit vector that is invariant ({\it i.e.} has 
   zero Lie derivative) under translations $\frac{\partial }{\partial x}$,  
   $\frac{\partial }{\partial y}$ in the transverse plane, boosts 
   $\frac{\partial}{\partial\varsigma}$, and rotations 
   $\frac{\partial }{\partial \phi }$ around the beam axis ($z$ 
   direction). Therefore, the temperature and any other hydrodynamical   
   variables become functions of $\tau$ only.}
In order to obtain the velocity profile in Minkowski space we simply have to 
map back from the coordinate system
$\hat{x}^{\mu }=(\rho,\theta,\phi,\varsigma )$ to 
$x^{\mu}=(\tau ,r,\phi,\varsigma)$, combined with the appropriate Weyl 
rescaling of the fluid velocity \cite{Gubser:2010ui}: 
\begin{equation}
u_{\mu }=\tau \,\frac{\partial \hat{x}^{\nu }}{\partial x^{\mu }}\,\hat{u}%
_{\nu }\,.  \label{eq:velmap}
\end{equation}%
This results in the following expressions for the Milne
components of the fluid four-velocity $u_{\mu }$ in Minkowski space 
\cite{Gubser:2010ze,Gubser:2010ui}: 
\begin{equation}
u_{\tau }=-\cosh\kappa (\tau ,r)\,,\hspace{0.5cm}u_{r}=\sinh \kappa (\tau
,r)\,,\hspace{0.5cm}u_{\phi }=u_{\varsigma }=0\,,
\end{equation}
with the transverse flow rapidity
\begin{equation}
\kappa (\tau ,r)=\atanh \left( \frac{2q^{2}\tau r}
                                                    {1{+}q^{2}\tau ^{2}{+}q^{2}r^{2}}\right) \,.
\end{equation}
In de Sitter space the ideal hydrodynamic equations reduce to
a single continuity equation for the thermal equilibrium energy density 
$\hat{\varepsilon}$. For dissipative hydrodynamics one has to solve in addition an 
equation of motion for the shear-stress tensor $\hat{\pi}^{\mu\nu}$. 
Some aspects of the dissipative IS solution are discussed in Appendix 
\ref{app:IS}; for a more complete discussion we
direct the interested reader to Ref.~\cite{Marrochio:2013wla}. From 
the solutions for the hydrodynamical fields in de Sitter space one obtains 
the non-trivial solution in Minkowski space through the transformation 
rules \cite{Gubser:2010ze,Gubser:2010ui}
\begin{subequations}
\begin{align}
\varepsilon (\tau ,r)& =\frac{\hat{\varepsilon}(\rho (\tau ,r))}{\tau ^{4}}\,, 
\\
\pi _{\mu \nu }(\tau ,r)& = \frac{1}{\tau^2}
\frac{\partial \hat{x}^{\alpha}}{\partial x^{\mu }}
\frac{\partial \hat{x}^{\beta }}{\partial x^{\nu }}
\hat{\pi}_{\alpha \beta }(\rho (\tau ,r))\,.
\end{align}
%

\section{Relativistic Boltzmann equation in curved spaces}
\label{sec:Boltzeqn} 

The general relativistic Boltzmann equation for the \textit{on-shell}
one-particle distribution function is given by \cite{Cercignani,Debbasch1:2009,Debbasch2:2009} 
\end{subequations}
\begin{equation}
p^{\mu }\partial _{\mu }f+\Gamma _{\mu i}^{\lambda }p_{\lambda }p^{\mu}
\frac{\partial f}{\partial p_{i}}=\mathcal{C}[f]\,,  
\label{boltzmann1}
\end{equation}
where the distribution function $f=f(x^{\mu },p_{i})$ is defined in a
7-dimensional phase space. A point in this phase-space is
described by seven coordinates, the spacetime coordinates 
$x^{\mu }=(t,x,y,z)$ and the three spatial covariant momentum components 
$p_{i}=(p_{x},p_{y},p_{z})$. The zero component of the momentum is obtained 
from the on-shell condition $p_{0}=p_{0}(x^{\lambda },p_{i})$.
Moreover, in Eq.~\eqref{boltzmann1} the Christoffel symbol 
$\Gamma _{\mu i}^{\lambda }$ is defined as%
\footnote{\label{fn4}%
   For a general curved spacetime the Christoffel symbols are non vanishing 
   but they can also be nonzero for a given flat spacetime depending on the   
   choice of coordinates. For instance when parametrizing any vector 
   $x^{\mu }$ in Minkowski space by using Milne coordinates 
   $x^{\mu }=(\tau ,x,y,\varsigma )$, the Christoffel symbols with nonzero 
   components are $\Gamma _{\varsigma\varsigma }^{\tau }=\tau$ and
   $\Gamma _{\varsigma \tau }^{\varsigma}=
     \Gamma _{\tau \varsigma }^{\varsigma }=1/\tau$.}
\begin{equation}
\Gamma _{\mu \nu }^{\lambda }=\frac{g^{\lambda \beta }}{2}\left( \partial
_{\mu }g_{\beta \nu }+\partial _{\nu }g_{\beta \mu }-\partial _{\beta
}g_{\mu \nu }\right) \,.
\end{equation}%
Equation~\eqref{boltzmann1} is covariant under general coordinate
transformations $x^{\mu }\rightarrow \hat{x}^{\mu }(x^{\lambda })$, 
although not manifestly so \cite{Debbasch1:2009,Debbasch2:2009}.
There are other ways to write this equation where the general coordinate  covariance is explicit \cite{Debbasch2:2009}. For instance, one can define 
an off-shell distribution $F(x^{\mu },p_{\mu })$ that satisfies a
manifestly covariant Boltzmann equation 
\cite{Debbasch1:2009,Debbasch2:2009}. We will not make use of this approach since the form of Eq.~\eqref{boltzmann1} is more 
convenient for our purposes. The explicit form of the collision term 
on the right-hand side for $2{\,\leftrightarrow\,}2$ scattering can be found 
in \cite{Debbasch1:2009,Debbasch2:2009}. In this work, we restrict ourselves
to a simple approximation for the collisional kernel, the relaxation
time approximation (RTA), in which ${\mathcal{C}}[f]$ is given by
\cite{Anderson:1974,Bhatnagar:1954}
\begin{equation}
{\mathcal{C}}[f]=\frac{p{\cdot}u}{\tau _{\mathrm{rel}}}
\bigl[f(x^{\mu},p_{i}){-}f_{\mathrm{eq}}(x^{\mu },p_{i})\bigr].   
\label{eq:RTA}
\end{equation}
Here $u^{\mu }$ is the fluid velocity, $T$ is the temperature in the 
local rest frame, $\tau _{\mathrm{rel}}$ is the relaxation time which can 
depend on proper time, $f_{\mathrm{eq}}(x^{\mu},p_{i})=
f_{\mathrm{eq}}(p\cdot u/T)$ is the local equilibrium J\"{u}ttner 
distribution, and the fluid velocity is defined in the Landau frame. 

For a given distribution function $f(x^{\mu },p_{i})$ one obtains the
energy-momentum tensor $T^{\mu \nu }$ as the following 
moment of the distribution function \cite{DeGroot:1980dk,Cercignani}: 
\begin{equation}
T^{\mu \nu }(x)=\int \frac{d^{3}p}{(2\pi)^3\sqrt{-g}p^{0}}\, p^{\mu }p^{\nu }f(x^{\mu},p_{i})\,.  \label{eq:enemomten}
\end{equation}
The relevant macroscopic variables such as the energy
density, pressure, etc., are most easily identified by decomposing 
the four-momentum into temporal and spatial parts {\it in the local 
rest frame}, $p^{\mu}=-(u\cdot p)u^{\mu }+\Delta ^{\mu \nu }p_{\nu }$, where ${-}u^\mu u^\nu$ and $\Delta ^{\mu\nu}=g^{\mu \nu }+u^{\mu}u^{\nu }$ are the projectors parallel and orthogonal to 
$u^{\mu }$. Given this vector decomposition, the energy-momentum 
tensor \eqref{eq:enemomten} for a theory with vanishing bulk viscosity (as 
is the case in a conformal theory) can be written as 
\begin{equation}
T^{\mu \nu }(x)=\varepsilon (x)u^{\mu }u^{\nu }+\Delta ^{\mu \nu }\mathcal{P}%
(x)+\pi ^{\mu \nu }(x)\,,
\end{equation}
where $\varepsilon $ is the energy density, $\mathcal{P}$ is the
thermodynamic pressure, and $\pi ^{\mu \nu }$ is the shear-stress tensor
which is traceless, symmetric, and orthogonal to the fluid velocity. The
macroscopic quantities above can be obtained as momentum 
moments of an arbitrary distribution function \cite{Cercignani}: 
\begin{subequations}
\label{eq:macrquant}
\begin{align}
\varepsilon (x)& =\int \frac{d^{3}p}{(2\pi)^3\sqrt{-g}p^{0}}\,(p\cdot u)^{2}
f(x^{\mu},p_{i})\,,   
\\
\mathcal{P}(x)& =\frac{1}{3}\int \frac{d^{3}p}{(2\pi)^3\sqrt{-g}p^{0}}\,
\Delta _{\mu\nu }p^{\nu }p^{\mu }f(x^{\mu },p_{i})\,, 
\label{eq:totpres}\\
\pi ^{\mu \nu }(x)& =\int \frac{d^{3}p}{(2\pi)^3\sqrt{-g}p^{0}}\,
p^{\langle \mu}p^{\nu \rangle }f(x^{\mu },p_{i})\,.\label{eq:shear}
\end{align}
\end{subequations}
In Eq.~\eqref{eq:shear} we introduce the notation $p^{\langle \mu }p^{\nu
\rangle }{\,=\,}\Delta _{\alpha \beta }^{\mu \nu }p^{\alpha }p^{\beta }$ where the double projector
$\Delta _{\alpha \beta }^{\mu \nu }=\left( \Delta _{\alpha }^{\mu }\Delta
_{\beta }^{\nu }+\Delta _{\beta }^{\mu }\Delta _{\alpha }^{\nu }-\frac{2}{3}
\Delta ^{\mu \nu }\Delta _{\alpha \beta }\right) /2$ selects the traceless and orthogonal (to $u^\mu$) part of a tensor. We will make use of these relations in Sec.~\ref{sec:Gubsersol} to study the dynamics of the macroscopic variables for a distribution function that exactly 
solves the RTA Boltzmann equation with Gubser symmetry.

\subsection{The RTA Boltzmann equation in Milne coordinates}
\label{subsec:Milne} 

For a system with boost-invariant longitudinal expansion it is convenient to use Milne coordinates $x^\mu=(\tau,x,y,\varsigma)$ 
with metric $g_{\mu\nu}=\mathrm{diag}(-1,1,1,\tau^2)$. In this 
coordinate system the RTA Boltzmann equation~\eqref{boltzmann1} is written as 
(see footnote~\ref{fn4})
\begin{equation}
p^\tau \partial_\tau f + p_x \partial_x f+p_y \partial_y f 
+ \frac{p_\varsigma}{\tau^2}\partial_\varsigma f 
= \frac{ p\cdot u}{\tau_\mathrm{rel}} \bigl(f{-}f_{\mathrm{eq}}\bigr) .  
\label{RTAmilne}
\end{equation}
The on-shell condition allows us to determine $p^\tau$:
\begin{equation}
p_\mu p_\nu g^{\mu\nu}=-m^2 \Longrightarrow p^\tau = 
\sqrt{m^2+p_x^2+p_y^2+p_\varsigma^2/\tau^2}\,.  
\label{defineptau}
\end{equation}
Since the Gubser symmetry to be studied in the following Section 
embodies conformal invariance which would be broken by non-zero mass 
terms, we set all masses to zero in this work.

\subsection{Emergent Weyl invariance in the massless limit}
\label{subsec:Weyl} 

Building on recent work on Weyl invariant hydrodynamics 
\cite{Baier:2007ix} for systems close to local equilibrium, we here use 
similar techniques to study their non-equilibrium dynamics, as described 
by the Boltzmann equation. We start by showing that, for massless
particles, conformal transformations are a symmetry of the Boltzmann 
equation in the RTA approximation.

Under a Weyl transformation, the metric changes as
\begin{equation}
g_{\mu\nu}(x) \to e^{-2\Omega(x)}g_{\mu\nu}(x) \, ,
\end{equation}
where $\Omega(x)$ is an arbitrary scalar function. A Weyl rescaling is not a
general coordinate transformation: The Ricci scalar changes, so
a flat space transforms into a curved space.%
\footnote{\label{fn5}%
   A specific example was discussed in Sect.~\ref{sec:Gubser}: 
   Due to the global factor $1/\tau^2$, the measure $d\hat{s}^2$ in
   \eqref{metricdS3R} does not parametrize standard Minkowski space
   $R^3\otimes R$.}
This transformation is similar to the Mercator projection map used in
cartography which maps the surface of the earth to a plane. Under a 
Weyl rescaling, a $(m,n)$ tensor 
$Q^{\mu_1 \ldots \mu_m}_{\nu_1 \ldots \nu_n}$ transforms as follows
\cite{Baier:2007ix}:
\begin{equation}  
\label{eq:Weyltrans}
Q^{\mu_1 \ldots \mu_m}_{\nu_1 \ldots \nu_n}(x) \to
e^{(\Delta+m-n)\Omega(x)}Q^{\mu_1 \ldots \mu_m}_{\nu_1 \ldots \nu_n}(x) \, ,
\end{equation}
where $\Delta$ is its canonical dimension, $m$ is the number of
contravariant indices, and $n$ is the number of covariant indices. For
example, the velocity vector transforms as 
$u_\mu \to e^{-\Omega}u_\mu$, $u^\mu \to e^{\Omega}u^\mu$ while 
the temperature transforms as $T \to e^{\Omega}T$. Scalar products
of four-vectors transform with the sum of their canonical dimensions.

Note that, to prove the Weyl covariance of the Boltzmann equation
\eqref{boltzmann1} we note that, consistent with its interpretation as a
probability density in phase-space, the distribution function 
$f(x^{\mu },p_{i})$ in Eq.~\eqref{boltzmann1} is a scalar with zero 
canonical dimension and thus invariant under Weyl transformations. Since 
for a conformal system $\tau_\mathrm{rel} \sim 1/T$, the RTA collision 
term \eqref{eq:RTA} on the r.h.s. of the Boltzmann equation 
\eqref{boltzmann1} transforms homogeneously as 
$\mathcal{C}[f]\to e^{2\Omega}\,\mathcal{C}[f]$ (as argued in 
\cite{Baier:2007ix} for a general collision term). Given \eqref{eq:Weyltrans},
the first term on the l.h.s. of Eq.~\eqref{boltzmann1} and the 
term multiplied by the Christoffel symbol are seen to transform in the same 
way: $p^\mu \partial_\mu f \to e^{2\Omega} p^\mu \partial_\mu f$ 
and $p_{\lambda }p^{\mu}\frac{\partial f}{\partial p_{i}}
\to e^{2\Omega} p_{\lambda }p^{\mu} \frac{\partial f}{\partial p_{i}}$. 
However, the Christoffel symbol $\Gamma_{\mu i}^\lambda$ itself transforms
non-trivially under a Weyl rescaling: 
\begin{equation}
\Gamma_{\mu\nu}^\lambda \to \Gamma_{\mu\nu}^\lambda-\left(g_\nu^\lambda
\partial_\mu \Omega+g_{\mu}^\lambda \partial_\nu
\Omega-g_{\mu\nu}\partial^\lambda \Omega\right).  
\label{changegamma}
\end{equation}
This leads to an additional term on the l.h.s. of the Boltzmann equation \eqref{boltzmann1}:
\begin{eqnarray}
\left(g_i^\lambda \partial_\mu \Omega+g_{\mu}^\lambda \partial_i
\Omega-g_{\mu i}\partial^\lambda \Omega\right) p_\lambda p^\mu 
\frac{\partial f}{\partial p_i} 
&=& \left(p_i\, p{\cdot}\partial\Omega+\partial_i\Omega\, p{\cdot}p
        -p_i\,p{\cdot}\partial \Omega\right) \frac{\partial f}{\partial p_i}  
\notag \\
&=& p{\cdot}p\,\partial_i\Omega\,\frac{\partial f}{\partial p_i} \, .
\end{eqnarray}
This term vanishes identically if and only if the particles are 
massless, {\it i.e.} if $p{\cdot}p{\,=\,}0$. Under these conditions one sees 
that the entire Boltzmann equation transforms homogeneously 
with $e^{2\Omega}$ under Weyl rescaling: 
\begin{equation}
p^\mu \partial_\mu f + \Gamma^{\lambda}_{\mu i}p_\lambda p^\mu 
\frac{\partial f}{\partial p_i} - \mathcal{C}[f]=0 \Longrightarrow
e^{2\Omega}\left(p^\mu \partial_\mu f + \Gamma^{\lambda}_{\mu i}p_\lambda
p^\mu \frac{\partial f}{\partial p_i} - \mathcal{C}[f]\right)=0\,.
\label{boltzmann1weyl}
\end{equation}
In the massless limit, the microscopic dynamics encoded in the 
distribution function thus possesses an ``emergent" Weyl
invariance, and conformal transformations from
Minkowski to curved spaces are then indeed symmetries of the Boltzmann 
equation. In the following Section we will obtain an exact solution 
of the RTA Boltzmann equation by exploiting the conformal map 
between Minkowski space and $dS_3\otimes R$.

\subsection{Solving the Boltzmann equation with Bjorken symmetry}
\label{sec:Bjo} 

Before doing so we would like to return to the case of boost invariant
longitudinal expansion with translational and rotational symmetry in the
transverse plane studied by Bjorken \cite{Bjorken:1982qr}. Transverse homogeneity implies that the distribution function $f(x^\mu,p_i)$ cannot 
depend on the transverse coordinates $x$ and $y$, while azimuthal 
symmetry around the longitudinal axis stipulates that any dependence on 
the transverse momentum components can be only through 
$p_T=\sqrt{p_x^2{+}p_y^2}$. Longitudinal boost invariance implies 
that any time dependence of $f$ can only occur in terms of the longitudinal
proper time ${\tau\,=\,}\sqrt{t^2{-}z^2}$, and that any dependence on the longitudinal position $z$ and the longitudinal momentum component $p_z$
must come in the boost invariant combination%
\footnote{\label{fn6}%
   The Lorentz $\gamma$ factor for the Bjorken flow profile    
   $v_z{\,=\,}z/t$ is $\gamma{\,=\,}(1{-}v_z^2)^{-1/2}=t/\tau$. 
   A particle with momentum $p_z$ at position $z$ in the lab frame thus has 
   momentum 
   $p^{\prime}_z=\gamma(p_z-E v)=(tp_z-Ez)/\tau\equiv w/\tau$ in the
   local rest frame. Since the physics in the local rest frame, in particular 
   the $p'_z$ distribution, is supposed to be independent of $z$, $f$ can   
   depend on $z$ and $p_z$ only through $w$.}   
\begin{equation}  
\label{eq:omega}
w = t p_z - z E\,.
\end{equation}
We see that longitudinal boost invariance imposes strong constraints
on the number of independent variables of the distribution function and on the particular combination in which the dependent variables appear 
\cite{Baym:1984np,Baym:1985tna,Bialas:1984wv,Bialas:1987en,%
Florkowski:2013lza,Florkowski:2013lya,Florkowski:2014sfa}. As we shall see in the following section, identifying these independent combinations
of phase-space variables will be the key step in deriving the exact solution 
of the Boltzmann equation for an $SO(3)_q\otimes SO(1,1)\otimes Z_2$ symmetric system.

With the above simplifications the RTA Boltzmann equation 
\eqref{boltzmann1} reduces to \cite{Baym:1984np,%
Florkowski:2013lza,Florkowski:2013lya,Florkowski:2014sfa} 
\begin{equation}
\partial_\tau f = -\frac{1}{\tau_{\mathrm{rel}}(\tau)}
\bigl(f-f_{\mathrm{eq}}\bigr)
\label{mikeeq}
\end{equation}
where $f=f(\tau;p_T,w)$. Its general solution for a momentum-independent
relaxation time $\tau_\mathrm{rel}$ is given by 
\begin{equation}  
\label{eq:RTAsol}
f(\tau;p_T,w)= D(\tau,\tau_0)f_0(p_T,w)+\int_{\tau_0}^\tau\,
\frac{d\tau^{\prime }}{\tau_{\mathrm{rel}}(\tau^{\prime })}\,
D(\tau,\tau^{\prime})\,f_{\mathrm{eq}}(\tau^{\prime};p_T,w) \, ,
\end{equation}
where $D(\tau_2,\tau_1)$ is the damping function given by 
\begin{equation}
D(\tau_2,\tau_1)=\exp\!\left(-\int_{\tau_1}^{\tau_2}\,\frac{d\tau''}{\tau_{\mathrm{rel}}(\tau'')}\right)\,,
\end{equation}
and $f_0(p_T,w)$ is the initial distribution function at $\tau{\,=\,}\tau_0$ \cite{Baym:1984np,Baym:1985tna,Bialas:1984wv,%
Bialas:1987en,Florkowski:2013lza,Florkowski:2013lya,Florkowski:2014sfa}. 
In the past, different authors considered an equilibrium initial condition 
$f_0(p_T,w)=f_{\mathrm{eq}}(\tau_0;p_T,w)$ \cite{Baym:1984np,%
Baym:1985tna,Bialas:1984wv,Bialas:1987en}. Recently Florkowski {\it et al.}
\cite{Florkowski:2013lza,Florkowski:2013lya} relaxed this assumption and 
studied a more general set of initial profiles for $f_0$, corresponding to 
an initially anisotropic local momentum distribution. The approach 
of Florkowski~{\it et al.} has proven very useful since it allows one to test 
different viscous and anisotropic hydrodynamic approximation schemes
\cite{Martinez:2010sc,Florkowski:2010cf,Ryblewski:2010bs,Martinez:2010sd,%
Ryblewski:2011aq,Florkowski:2011jg,Martinez:2012tu,Ryblewski:2012rr,%
Florkowski:2012as,Bazow:2013ifa,Florkowski:2013uqa,Tinti:2013vba,%
Florkowski:2014txa,Florkowski:2014bba} against the underlying microscopic
Boltzmann dynamics, for both massless 
\cite{Florkowski:2013lza,Florkowski:2013lya} 
and massive cases \cite{Florkowski:2014sfa}.%
\footnote{\label{fn7}%
   We point out that the solution \eqref{eq:RTAsol} for the distribution 
   function was derived using only the symmetry constraints,
   without assumptions about the particle mass    
   \cite{Baym:1984np}. This will be different for the case of Gubser symmetry
   studied in the following Section.}

\section{Exact solution of the RTA Boltzmann equation with Gubser symmetry}
\label{sec:Gubsersol}
\subsection{The solution}
\label{sec4a} 

We now discuss the consequences for the microscopic kinetic evolution of the system of requiring invariance under
$SO(3)_{q}\otimes SO(1,1)\otimes Z_{2}$ transformations. The previous 
Section taught us that this is most easily studied in a coordinate system 
where the fluid is at rest. In addition, one should use a spacetime where 
the group symmetries are explicitly manifest. Hence, as pointed out in
Sec.~\ref{sec:Gubser}, the most natural choice is to use the space $dS_{3}\otimes R$ parametrized by coordinates
$\hat{x}^{\mu }=(\rho ,\theta ,\phi ,\varsigma )$.

As in the case of Bjorken symmetry, the Gubser symmetry severely
restricts the combinations of the coordinates 
$\hat{x}^\mu=(\rho,\theta,\phi,\varsigma)$ and momenta 
$\hat{p}_i=(\hat{p}_\theta,\hat{p}_\phi,\hat{p}_\varsigma)$ on which 
$f(\hat{x}^\mu,\hat{p}_i)$ can depend. Each of the three factors of the 
Gubser group $SO(3)_q\otimes SO(1,1)\otimes Z_2$ imposes its own constraints:
\begin{itemize}
\item The generators of $SO(3)_q$ describe rotations of spatial vectors 
parametrized by the $(\theta,\phi)$ variables, and simultaneously of momentum 
vectors parametrized by the coordinates $(\hat{p}_\theta,\hat{p}_\phi)$, over a 
sphere $S^2$. $SO(3)_q$ symmetry demands that physical observables (such as the 
distribution function $f$) can only depend on the following $SO(3)_q$-invariant 
combination of these four variables:
\begin{equation}  
\label{eq:pomega}
\hat{p}_\Omega^2=\hat{p}_\theta^2 + \frac{\hat{p}_\phi^2}{\sin^2\theta}\,.
\end{equation}
Geometrically $\hat{p}_\Omega^2$ is the radius of the sphere $S^2$
in $(\hat{p}_\theta,\hat{p}_\phi)$ momentum-space coordinates.%
\footnote{\label{fn8}%
   This situation is analogous to what happens in the hydrogen atom problem in
   quantum mechanics. There, the only combination of the generators
   $L_x$, $L_y$ and $L_z$ of the $SO(3)$ angular momentum algebra with which
   all three generators commute is the Casimir operator 
   $L^2=L_x^2+L_y^2+L_z^2$.}
Thus, we interpret $\hat{p}_\Omega$ as the total momentum 
associated with the momentum components $(\hat{p}_\theta,\hat{p}_\phi)$. 

\item The $SO(1,1)$ invariance imposes the same constraints on the
distribution function as previously discussed in Sec.~\ref{sec:Bjo} for the 
case of Bjorken symmetry: in Milne coordinates, the distribution function 
depends only on the proper time $\tau$, the transverse momentum $p_{T}$ and
the variable $w$ \eqref{eq:omega}, but not on the spatial rapidity 
$\varsigma$. The $SO(3)_{q}$ symmetry modifies the dependence on 
the first two variables ($\tau$ and $p_{T}$) in de Sitter space. Given the 
conformal map between Minkowski and de Sitter space one can show 
that the variable $w$ is related to the $\hat{p}_\varsigma$ component.%
\footnote{\label{fn9}%
   The covariant components of the momentum transform
   as \cite{Cercignani}
   \begin{equation}
   p^{\prime \mu }=\frac{\partial x^{\prime \mu }}{\partial x^{\nu }}\,
   p^{\nu}\,.
   \end{equation}
   When transforming from $(t,x,y,z)$ to $(\tau ,x,y,\varsigma )$
   coordinates, this prescription yields for the 
   $\varsigma$-component of the momentum the expression
   \begin{equation*}
   p^\varsigma=\frac{1}{\tau^2}(tp^z-zp^0)=\frac{w}{\tau^2}=\frac{p_\varsigma}{\tau^2}\,,
   \end{equation*}
   where we used the definition \eqref{eq:omega} of the 
   variable $w$. Under Weyl rescaling (\ref{eq:velmap}) this component
   transforms into 
   $\hat{p}^\varsigma=\tau^2 p^\varsigma= p_\varsigma=\hat{p}_\varsigma=w$. Note that
   under a boost with rapidity $\varsigma _{boost}$, 
   $\varsigma \to \varsigma+\varsigma _{boost}$ while 
   $\hat{p}^{\varsigma }$ remains invariant.}
Therefore, the $SO(1,1)$ invariance implies that, in addition to  
$\hat{p}_\Omega$, the distribution function can depend in momentum space 
only on $\hat{p}_\varsigma$. 

\item The $Z_2$ invariance implies that the distribution function is invariant
under reflection $\varsigma\to -\varsigma$.
\end{itemize}

As a result of these considerations we see that the conformal symmetry 
group demands that 
$f(\hat{x}^\mu,\hat{p}_i)=f(\rho;\hat{p}^2_\Omega,\hat{p}^\varsigma)$. Its
only dependence on the spacetime coordinates is through the ``de Sitter time'' 
$\rho$. Conformality also imposes constraints on the functional dependence of the 
relaxation time on the temperature: $\tau_{\mathrm{rel}}=c/T$ where $c$ is a 
free dimensionless parameter related to the shear viscosity to entropy density ratio $\eta/{\mathcal{S}}$. For the RTA collision kernel used in this work
one has \cite{Denicol:2010xn,Denicol:2011fa,Florkowski:2013lza,Florkowski:2013lya}
\begin{equation}
\label{eq:c}
c = \frac{5 \eta}{\mathcal{S}} \ 
\Longleftrightarrow\ 
\frac{\eta}{\mathcal{S}} = \frac{1}{5}\, \tau_\mathrm{rel} T\, ,
\end{equation}
where $\eta$ is the shear viscosity and $\mathcal{S}$ the entropy
density. Due to the conformal map between Minkowski space and 
$dS_3\otimes R$ one has the relation 
\begin{equation}
T(\tau,r)=\hat{T}(\rho(\tau,r))/\tau \, .  
\label{eq:That}
\end{equation}

Putting all these ingredients together, Gubser invariance is seen to 
greatly simplify the RTA Boltzmann equation \eqref{RTAmilne}. Starting with the RTA Boltzmann equation in Milne coordinates, changing the variables from $x^\mu \to \hat{x}^\mu$ and performing the necessary Weyl rescalings, one finds that the kinetic equation can be written in $dS_3\otimes R$ coordinates as
\begin{equation}
\frac{\partial}{\partial \rho}f(\rho;\hat{p}_\Omega^2,\hat{p}_\varsigma)
= -\frac{\hat{T}(\rho)}{c}\left[ f\!
\left(\rho;\hat{p}_\Omega^2,\hat{p}_\varsigma\right)
- f_{\mathrm{eq}}\!\left(\hat{p}^\rho/\hat{T}(\rho)\right) %
\right] ,  
\label{newRTAboltzmanneq}
\end{equation}
where $\hat{p}^\rho$ is determined from the massless on-shell condition
\begin{equation}  
\label{eq:prho}
\hat{p}^\rho =\sqrt{\frac{\hat{p}_\Omega^2}{\cosh^2\!\rho}+\hat{p}%
_\varsigma^2}\,.
\end{equation}
Comparing Eqs.~\eqref{newRTAboltzmanneq} and \eqref{mikeeq}, 
the solution for $f(\rho;\hat{p}^2_\Omega,\hat{p}_\varsigma)$ is easily found:
\begin{equation}
f(\rho;\hat{p}^2_\Omega,\hat{p}_\varsigma)=
D(\rho,\rho_0) f_0(\rho_0;\hat p_\Omega^2,\hat{p}_\varsigma)
+\frac{1}{c}\int_{\rho_0}^\rho d\rho^{\prime}\,D(\rho,\rho^{\prime })\,
\hat T(\rho^{\prime })\, 
f_{\mathrm{eq}}(\rho^{\prime};\hat p_\Omega^2,\hat{p}_\varsigma) \, .
\label{boltzmannsolution}
\end{equation}
The damping function in this case is given by 
\begin{equation}
D(\rho_2,\rho_1)=\exp\!\left(-\int_{\rho_1}^{\rho_2} d\rho''\,
\frac{\hat T(\rho'')}{c} \right) .  
\label{defineD}
\end{equation}
In Eqs.~\eqref{boltzmannsolution} and~\eqref{defineD}, $\rho_0$ is the
initial ``time" in de Sitter space at which 
$f=f_0(\rho_0;\hat p_\Omega^2,\hat{p}_\varsigma)$. In the present work 
we assume that $f_0$ is given by a Boltzmann
equilibrium distribution function at $\rho_0$. With this assumption the Weyl rescaling
property is preserved for any value of the $\rho$ variable. 

\subsection{Energy-momentum tensor components}
\label{subsec:enemomcomp} 

The solution \eqref{boltzmannsolution} of the Boltzmann equation allows
us to calculate the evolution of all components of the energy-momentum tensor 
from their definitions, Eqs.~\eqref{eq:macrquant}. For reference, let us
first calculate the energy density for a Boltzmann equilibrium
distribution function:
\begin{equation}
\label{eq:eqenedens}
\begin{split}
{\hat{\varepsilon}_\mathrm{eq}}(\rho )& =
\frac{1}{(2\pi )^{3}}\int_{-\infty }^{\infty }d\hat{p}_{\varsigma}\,
\int_{-\infty }^{\infty }\frac{d\hat{p}_{\theta }}{\cosh \rho }\,
\int_{-\infty }^{\infty }\frac{d\hat{p}_{\phi }}{\cosh\rho\,\sin\theta}\,
\hat{p}^{\rho }\, e^{-\hat{p}^{\rho }/\hat{T}(\rho )} 
\\
& =\frac{3}{\pi ^{2}}\hat{T}^{4}(\rho )\,.
\end{split}
\end{equation}
As expected for a conformal theory, 
$\hat{\varepsilon}_\mathrm{eq}\sim \hat{T}^{4}$. The energy density 
associated with the exact solution \eqref{boltzmannsolution} of the Boltzmann 
equation is obtained as follows:
\begin{equation}
\begin{split}
{\hat{\varepsilon}}(\rho )& =\frac{1}{(2\pi )^{3}}
\int_{-\infty }^{\infty } d\hat{p}_{\varsigma }\,
\int_{-\infty }^{\infty }\frac{d\hat{p}_{\theta }}{\cosh\rho}\,
\int_{-\infty }^{\infty }\frac{d\hat{p}_{\phi }}{\cosh\rho\,\sin\theta}\,
\hat{p}^{\rho }\, f(\rho;\hat{p}_{\Omega }^{2},\hat{p}_\varsigma)
\\
& =\frac{3}{\pi ^{2}}\left[ D(\rho ,\rho _{0})\mathcal{H}\!\left( \frac{%
\cosh \rho _{0}}{\cosh \rho }\right) \hat{T}_{0}^{4}+\frac{1}{c}\int_{\rho
_{0}}^{\rho }d\rho ^{\prime }\,D(\rho ,\rho ^{\prime })\,\mathcal{H}\!\left( 
\frac{\cosh \rho ^{\prime }}{\cosh \rho }\right) \,\hat{T}^{5}(\rho ^{\prime
})\right] ,
\end{split}
\label{eq:exactene}
\end{equation}
where in the last line we used both 
Eqs.~\eqref{boltzmannsolution} and \eqref{eq:eqenedens}. In 
Eq.~\eqref{eq:exactene} ${\hat{T}}_{0}\equiv {\hat{T}}(\rho _{0})$, and the 
function $\mathcal{H}(x)$ is 
\begin{equation}
\mathcal{H}(x)=\frac{1}{2}\left( x^{2}+x^{4}\frac{\tanh ^{-1}
\left( \sqrt{1{-}x^{2}}\right) }{\sqrt{1{-}x^{2}}}\right) .  
\label{eq:calh}
\end{equation}
It is straightforward to show that for the distribution function
\eqref{boltzmannsolution} the pressure \eqref{eq:totpres} is 
related to the energy density by the conformal equation of state
$\hat{\mathcal{P}}(\rho){\,=\,}\hat{\varepsilon}(\rho)/3$ at all
de Sitter times $\rho$.

From its definition, the shear stress tensor \eqref{eq:shear} is 
\begin{equation}
\hat{\pi}^{\mu \nu }=\frac{1}{(2\pi )^{3}}
\int_{-\infty }^{\infty }d\hat{p}_{\varsigma }\,
\int_{-\infty }^{\infty }\frac{d\hat{p}_{\theta}}{\cosh\rho}\,
\int_{-\infty }^{\infty }\frac{d\hat{p}_{\phi }}{\cosh\rho\,\sin\theta}\,
\frac{1}{\hat{p}^{\rho }}\hat{p}^{\langle\mu}\hat{p}^{\,\nu \rangle}\,
f(\rho;\hat{p}_{\Omega }^{2},\hat{p}_\varsigma)  
\label{definepi}
\end{equation}

In the $(\rho ,\theta ,\phi ,\varsigma )$ coordinate system the only nonzero
components of the shear stress tensor are 
\begin{subequations}
\label{eq:shearcomp}
\begin{align}
\hat{\pi}^\varsigma_\varsigma (\rho)& =
\frac{1}{(2\pi )^{3}}\int_{-\infty}^{\infty }d\hat{p}_{\varsigma }
\int_{-\infty }^{\infty }\frac{d\hat{p}_{\theta }}{\cosh^2\rho}
\int_{-\infty }^{\infty }\frac{d\hat{p}_{\phi }}{\sin \theta }\,
\frac{1}{\hat{p}^{\rho }}\,\left( \hat{p}_{\varsigma }^{2}-%
\frac{(\hat{p}^{\rho })^{2}}{3}\right) \!f  
\notag  \\
& =\frac{1}{\pi ^{2}}\left[ D(\rho ,\rho _{0})\mathcal{A}\!
\left( \frac{\cosh\rho}{\cosh\rho_{0}}\right) \hat{T}_{0}^{4}
+\frac{1}{c}\int_{\rho_{0}}^{\rho }d\rho ^{\prime }\,D(\rho ,\rho ^{\prime })\,\mathcal{A}\!\left(\frac{\cosh\rho}{\cosh\rho^{\prime }}\right) 
\hat{T}^{5}(\rho ^{\prime})\,\right], 
 \label{eq:varvar} \\
\hat{\pi}^\theta_\theta(\rho) & =\frac{1}{(2\pi )^{3}}
\int_{-\infty }^{\infty }d\hat{p}_{\varsigma }
\int_{-\infty }^{\infty }\frac{d\hat{p}_{\theta }}{\cosh^2\rho}
\int_{-\infty }^{\infty }\frac{d\hat{p}_{\phi }}{\sin\theta}\,
\frac{1}{\hat{p}^{\rho }}\,\left(\frac{\hat{p}_{\theta}^{2}}{\cosh^{2}\rho} 
-\frac{(\hat{p}^{\rho })^{2}}{3}\right) \!f 
= - \frac{1}{2}\hat{\pi}^\varsigma_\varsigma(\rho)\,,
\label{eq:thetatheta}\\
\hat{\pi}^\phi_\phi(\rho) & =
\frac{1}{(2\pi )^{3}}\int_{-\infty }^{\infty }d\hat{p}_{\varsigma }
\int_{-\infty}^{\infty }\frac{d\hat{p}_{\theta }}{\cosh^2\rho}
\int_{-\infty}^{\infty }\frac{d\hat{p}_{\phi }}{\sin \theta }\,
\frac{1}{\hat{p}^{\rho }}\,\left( \frac{\hat{p}_{\phi }^{2}}
{\cosh ^{2}\rho \sin ^{2}\theta }
-\frac{(\hat{p}^{\rho })^{2}}{3}\right) \!f = - \frac{1}{2}\hat{\pi}^\varsigma_\varsigma(\rho)\,.
\label{eq:phiphi} 
\end{align}
\end{subequations}
Here $f=f(\rho;\,\hat{p}_{\Omega }^{2},\hat{p}_\varsigma)$, and 
we defined 
\begin{equation}
\label{eq:FKGfunc}
\mathcal{A}(x) =\frac{x\sqrt{x^2{-}1}(1{+}2x^{2})
+(1{-}4x^2)\acoth\bigl(x/\sqrt{x^2{-}1}\bigr)}{2x^3(x^2{-}1)^{3/2}}\,. 
\end{equation}
Clearly, these expressions are consistent with $SO(3)_{q}$ symmetry
which demands $\hat{\pi}_{\theta }^{\theta }=\hat{\pi}_{\phi }^{\phi}$, and with 
the tracelessness of the shear stress tensor, 
$\hat{\pi}_{\theta }^{\theta}+\hat{\pi}_{\phi}^{\phi}
+\hat{\pi}_{\varsigma }^{\varsigma }=0$. Eqs.~\eqref{eq:shearcomp} tell us that  
for a fluid undergoing Gubser flow 
\cite{Gubser:2010ze,Gubser:2010ui,Marrochio:2013wla} the shear stress tensor 
has only a single independent non-zero component for which we choose
$\hat{\pi}^\varsigma_\varsigma$. One checks easily that, as expected, all 
shear stress components in \eqref{eq:shearcomp} approach zero when 
$f \to f_{\mathrm{eq}}$.

\subsection{Matching condition: definition of temperature}
\label{subsec:matching} 

For any point $\rho$, we define the local temperature 
$\hat{T}(\rho)$ of the fluid using the traditional matching condition 
\begin{equation}
\hat{\varepsilon}(\rho ) = 
\hat{\varepsilon}_{\mathrm{eq}}\bigl(\hat{T}(\rho)\bigr)
= \frac{3}{\pi^2}\hat{T}^4(\rho).
\label{eq:matchcond}
\end{equation}
Inserting this into Eq.~\eqref{eq:exactene}  we obtain the following integral equation for the temperature of the system:
\begin{equation}
\hat{T}^{4}(\rho )=D(\rho ,\rho _{0})\mathcal{H}
\left( \frac{\cosh \rho _{0}}{\cosh\rho}\right) \hat{T}^{4}(\rho _0)
+\frac{1}{c}\int_{\rho_0}^{\rho}d\rho'\,D(\rho,\rho')\,
\mathcal{H}\left(\frac{\cosh\rho'}{\cosh\rho}\right) \,\hat{T}^{5}(\rho')\,.  
\label{eq:efftemp}
\end{equation}
This integral equation can be solved iteratively. Once $\hat{T}(\rho)$ 
has been  determined, it can be used to calculate from
Eq.~(\ref{boltzmannsolution}) the full distribution function and from 
Eq.~\eqref{eq:varvar} the non-vanishing component of the shear
stress tensor.

\subsection{Conformal hydrodynamics from the exact kinetic solution}
\label{subsec:confhydro} 

In this subsection we show how to obtain first-
\cite{Gubser:2010ze,Gubser:2010ui} and second-order \cite{Marrochio:2013wla} 
conformal viscous hydrodynamics from the exact solution 
\eqref{boltzmannsolution} of the Boltzmann equation. To obtain the evolution equation for the viscous shear stress tensor one can follow the method described in \cite{Denicol:2010xn,Denicol:2012cn}. Our 
starting point is to rewrite the Boltzmann equation~\eqref{newRTAboltzmanneq} in the following form 
\begin{equation}  
\label{eq:RTAdelta}
\frac{\partial\delta f}{\partial\rho}=
-\frac{\delta f}{\hat{\tau}_{\mathrm{rel}}(\rho)}
-\frac{\partial f_{\mathrm{eq}}}{\partial\rho} \, ,
\end{equation}
where $\hat{\tau}_{\mathrm{rel}}(\rho)=\tau_\mathrm{rel}(\rho)/\tau%
=c/\hat{T}(\rho)$ and $\delta f = f{-}f_{\mathrm{eq}}$. The general expression for the shear-stress tensor \eqref{eq:shear} in terms of $\delta f$ is 
\begin{equation}  
\label{eq:newpi}
\hat{\pi}^{\langle\mu\nu\rangle}=\frac{1}{(2\pi)^3}\int\frac{d^3\hat{p}}{%
\sqrt{-g}\hat{p}^\rho}\,\hat{\Delta}^{\mu\nu}_{\alpha\beta}\hat{p}^\alpha%
\hat{p}^\beta\,\delta f\,.
\end{equation}
Taking the derivative with respect to $\rho$ we obtain the equation of 
motion
\begin{equation}
\begin{split}
\partial_\rho\hat{\pi}^{\langle \mu\nu\rangle}=
\hat{\Delta}^{\mu\nu}_{\alpha\beta}\frac{\partial}
{\partial\rho}\hat{\pi}^{\alpha\beta}
&=\int\frac{d^3\hat{p}}{(2\pi)^3}\,\hat{\Delta}^{\mu\nu}_{\alpha\beta}\,
\hat{p}^\alpha\hat{p}^\beta\,\left[\frac{1}{\sqrt{-g} \hat{p}^\rho}
\frac{\partial\delta f}{\partial\rho}+\delta f\frac{\partial}{\partial\rho}
\left(\frac{1}{\sqrt{-g}\hat{p}^\rho}\right)\right] ,
\end{split}
\end{equation}
where $\hat{\Delta}^{\mu\nu}_{\alpha\beta}$ is the transverse and
traceless double projector defined in $dS_3\otimes R$ with 
$(\rho,\theta,\phi,\varsigma)$ coordinates. In this step all terms that
vanish in the massless limit have been dropped. Using the Boltzmann equation
\eqref{eq:RTAdelta} for $\delta f$ together with 
$f_{\mathrm{eq}}=e^{-\hat{p^{\rho }}/\hat{T}(\rho )}$ one obtains the following 
(exact but implicit) evolution equation: 
\begin{equation}
\begin{split}
\partial _{\rho }\hat{\pi}^{\langle \mu \nu \rangle }& =
-\frac{\hat{\pi}^{\mu \nu }}{\hat{\tau}_{\mathrm{rel}}}
-2\hat{\pi}^{\mu \nu }\tanh \rho  
\\
& -\frac{\tanh \rho }{\hat{T}}\int\frac{d^{3}\hat{p}}{(2\pi)^3}
\frac{\hat{\Delta}_{\alpha \beta }^{\mu \nu }\hat{p}^{\alpha }\hat{p}^{\beta }}
{\sqrt{-g}(\hat{p}^{\rho})^2}
\left( \frac{p_{\theta }^{2}}{\cosh ^{2}\rho }+\frac{p_{\phi }^{2}}
{\cosh^2\rho\,\sin^2\theta}\right) e^{-\hat{p}^\rho/\hat{T}(\rho)} 
\\
& -\int \frac{d^{3}\hat{p}}{(2\pi)^3}
\frac{\hat{\Delta}_{\alpha \beta}^{\mu \nu }\hat{p}^{\alpha }\hat{p}^{\beta }}{\sqrt{-g}\hat{p}^{\rho }}\,\delta f\,\frac{1}{\hat{p}^{\rho }}
\frac{\partial \hat{p}^{\rho }}{\partial\rho }\,.
\label{eq:47}
\end{split}%
\end{equation}
As already discussed, we need to work this out only for the $\hat{\pi}^{\varsigma\varsigma }$ component. Performing
the integral in the second line for $\mu=\nu=\varsigma$ one obtains, after 
some algebra, its evolution equation in the following form:
\begin{equation}
\label{eq:pikinetic}
\begin{split}
\frac{\partial }{\partial \rho }\hat{\pi}^{\varsigma \varsigma } =&
-\frac{\hat{\pi}^{\varsigma \varsigma }}{\hat{\tau}_{\mathrm{rel}}}
+\frac{4}{3}\frac{\hat{\eta}}{\hat{\tau}_{\mathrm{rel}}}\tanh \rho 
-\frac{46}{21}\hat{\pi}^{\varsigma \varsigma }\tanh \rho  
\\
& +\frac{\tanh \rho }{3\,(2\pi )}\int_{0}^{\infty }d\hat{p}^{\rho}
\int_{0}^{2\pi }d\theta \,(\hat{p}^{\rho })^{3}\sin\theta
\left(\frac{25}{21}{-}\cos^2\theta \right) (3\cos^2\theta{-}1) \,\delta f\,.
\end{split}%
\end{equation}
Here the second term on the r.h.s. arises from the integral over 
$f_\mathrm{eq}$ in \eqref{eq:47} where we used Eq.~\eqref{eq:c}, the 
thermodynamic relation $\varepsilon+\mathcal{P}=\mathcal{S} T$, and the
definition $\hat{\eta}\equiv\eta\tau^3$. The second and last terms on the r.h.s.
of Eq.~\eqref{eq:47} were rearranged, using the definition \eqref{eq:newpi}, to
give the last two terms in Eq.~\eqref{eq:pikinetic}.

The first line of Eq.~\eqref{eq:pikinetic} gives the second order conformal
IS evolution equation for the independent shear viscous component (details of the 
derivation of the second order viscous hydrodynamical approximations are found in 
App.~\ref{app:IS}). The second line is a correction arising from the exact treatment 
of the distribution function. It is precisely this type of correction that can be missed 
when an approximate method is used to solve the Boltzmann equation. In Sec.~\ref{sec:result} we will study how large these deviations are by comparing the exact kinetic solution with predictions from ideal and different
variants of second-order viscous hydrodynamics that were obtained from the Boltzmann equation using different approximation schemes. Finally, we note that the first-order NS 
solution is easily extracted from Eq.~\eqref{eq:pikinetic} by taking the limit 
$\hat{\tau}_{\mathrm{rel}}\rightarrow 0$: 
\begin{equation}
\hat{\pi}_{NS}^{\varsigma \varsigma }=\frac{4}{3}\eta \,\tanh \rho \,.
\end{equation}
This is precisely the exact solution to conformal NS theory previously obtained 
in \cite{Gubser:2010ze,Gubser:2010ui}.

\section{Results and discussion}
\label{sec:result} 

\begin{figure}[t]
\centering
\includegraphics[width=0.65\textwidth]{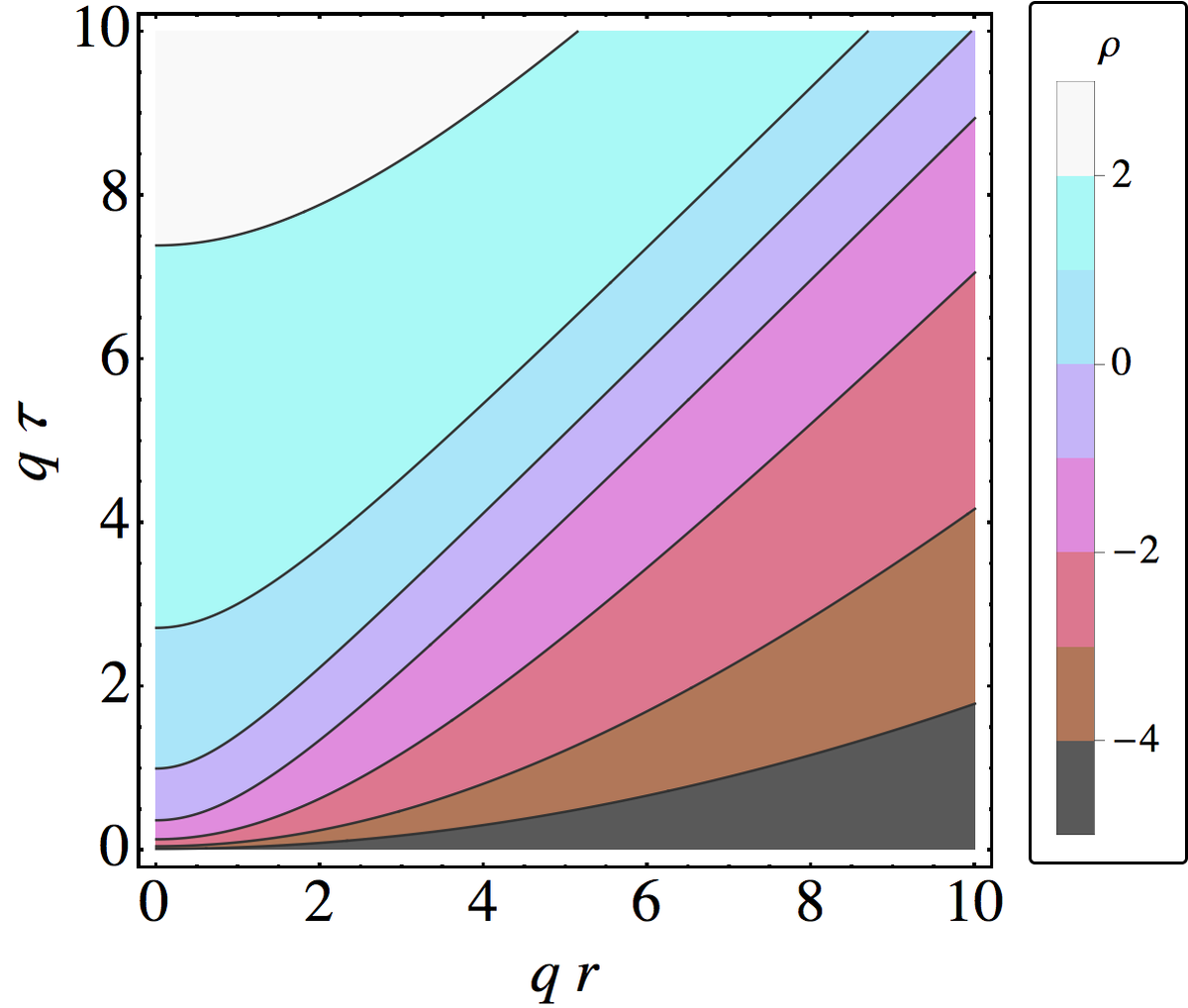}
\caption{(Color online) Lines of constant $\rho$ in the $(q\tau,qr)$ 
plane. The origin in de Sitter time, $\rho{\,=\,}0$, corresponds to the line 
going through $(q\tau,qr){\,=\,}(1,0)$ and the upper right corner of the 
graph.}
\label{F1}
\end{figure}

In this section we present solutions to Eq.~\eqref{boltzmannsolution}. To
obtain the solutions we first numerically solve the integral equation for
the effective temperature \eqref{eq:efftemp} using the method of iteration 
\cite{Florkowski:2013lza,Florkowski:2013lya,Florkowski:2014sfa}. One key
difference from the exact solutions obtained previously in 
\cite{Florkowski:2013lza,Florkowski:2013lya,Florkowski:2014sfa} is that one
must solve Eq.~\eqref{eq:efftemp} for both positive and negative values of 
the de Sitter time $\rho$. In addition, another key conceptual 
difference is that, instead of providing an initial condition as a function of the 
radius at fixed proper-time, we must instead specify an initial condition at a fixed
de Sitter time $\rho_0$ which maps to a line in $\tau$ and $r$ in Milne
coordinates (shown in Fig.~\ref{F1}).%
\footnote{\label{fn10}%
   Due to rotational symmetry and boost-invariance, by construction we can
   ignore the dependence on $\phi$ and $\varsigma$; however, in reality the
   surface is, in fact, three-dimensional. Note that, according 
   to Eq.~\eqref{eq:That}, a fixed temperature along a line of constant $\rho$ 
   implies a temperature profile that decreases like $1/\tau(r)$ as $r$ and 
   $\tau$ increase.}
Here we choose the initial condition at $\rho_0$ to be isotropic and ideal, 
{\it i.e.} we require that the shear-stress tensor vanishes at $\rho_0$. The 
code necessary to obtain the exact numerical solution is included as 
a supplemental file~\cite{weylRTAcode}.

For the numerical solution we discretize $\rho$ on an equally-spaced lattice
from $-\rho_{\mathrm{max}}$ to $+\rho_{\mathrm{max}}$ with 
$\rho_{\mathrm{max}} = 10$. The number of grid points required depends on 
$\eta/\mathcal{S}$. For $4\pi \eta/\mathcal{S}{\,=\,}0.1$ one needs on the order of
2000 grid points and on the order of 200 iterations; however, for larger 
$\eta/\mathcal{S}$ it is possible to use fewer grid points, and convergence
can be achieved in a much fewer number of iterations. For the initial guess
for the solution used in the iterations, we choose the ideal hydrodynamics
solution of Gubser and then iterate until the energy density converges to 
one part in $10^{10}$ at all points on the lattice. Once the 
effective temperature is obtained, it is used to compute the 
shear-stress, the full distribution function, and other observables, 
based on the results derived in the previous sections.

In what follows we will compare our numerical results with the free
streaming result and three hydrodynamical approximations: the ideal solution
of Gubser, the IS second-order viscous hydrodynamics solution of 
Ref.~\cite{Marrochio:2013wla}, and a (new) complete second-order viscous 
solution which we label as DNMR. The exact free streaming result, which 
corresponds to the limit $\eta/\mathcal{S} \rightarrow \infty$ ($c \to \infty$),
can be obtained for both the de Sitter space temperature profile and the $%
\varsigma\varsigma$ component of the shear-stress tensor using Eqs.~(\ref%
{eq:efftemp}) and (\ref{eq:varvar}), respectively. The results are
\begin{equation}
\hat{T}_{\mathrm{free\;streaming}} (\rho)= \mathcal{H}^{1/4}\!\left(\frac{%
\cosh\rho_0}{\cosh\rho}\right) \hat{T}_0(\rho_0) \, ,
\end{equation}
with $\mathcal{H}$ defined in Eq.~(\ref{eq:calh}), and 
\begin{equation}
\hat\pi^{\varsigma\varsigma}_{\mathrm{free\;streaming}}(\rho) = \mathcal{A}%
\!\left(\frac{\cosh\rho}{\cosh\rho_0}\right) \frac{\hat{T}_0^4}{\pi^2} \, ,
\end{equation}
with $\mathcal{A}$ defined in Eq.~(\ref{eq:FKGfunc}).

In the other limit $\eta/\mathcal{S} \rightarrow 0$ ($c \rightarrow 0$),
which corresponds to the ideal hydrodynamics case, one has~\cite%
{Gubser:2010ze,Gubser:2010ui} 
\begin{equation}
\label{eq:52}
\hat{T}_{\mathrm{ideal}}(\rho) = \frac{\hat{T}_0}{\cosh^{2/3}(\rho)} \, .
\end{equation}
For the second-order hydrodynamic approximation one has to 
solve two coupled ordinary differential equations subject to a boundary 
condition at $\rho{\,=\,}\rho_0$. For the IS case, the necessary equations are 
\begin{eqnarray}
&&\frac{1}{\hat{T}}\frac{d\hat{T}}{d\rho }+\frac{2}{3}\tanh \rho =
\frac{1}{3}\bar{\pi}_{\varsigma }^{\varsigma }(\rho )\,\tanh \rho \,, 
\label{eq:istemp}
\\
&&\frac{d\bar{\pi}_{\varsigma }^{\varsigma }}{d\rho }
+\frac{4}{3}\left( \bar{\pi}_{\varsigma }^{\varsigma }\right)^{2}\tanh \rho
+\frac{\bar{\pi}_\varsigma^\varsigma}{\hat{\tau}_\pi} =\frac{4}{15}\tanh\rho \,,
\label{eq:isshear}
\end{eqnarray}
where $\bar{\pi}_{\varsigma}^{\varsigma} \equiv 
\hat{\pi}_\varsigma^\varsigma/(\hat{T}\hat{\mathcal{S}})$ and 
$\hat{\tau}_\pi = 5\eta/(\mathcal{S}\hat{T})$. One can go beyond the 
IS approximation presented in Ref.~\cite{Marrochio:2013wla} and also include 
the complete second order contribution (see App.~\ref{app:IS} for further 
details). In this case, the second equation above should be replaced by
\begin{equation}
\frac{d\bar{\pi}_{\varsigma }^{\varsigma }}{d\rho }
+\frac{4}{3}\left(\bar{\pi}_{\varsigma }^{\varsigma }\right) ^{2}\tanh \rho 
+\frac{\bar{\pi}_\varsigma^\varsigma}{\hat{\tau}_\pi} 
= \frac{4}{15} \tanh \rho 
+ \frac{10}{21}\bar{\pi}_\varsigma^\varsigma \tanh\rho \,.
\label{eq:dnmrshear}
\end{equation}
If Eq.~(\ref{eq:dnmrshear}) is used, the result is labeled as DNMR.

For all cases shown in the results section we require as the boundary 
condition for the solution in de Sitter space that the system is ideal at 
$\rho_0=0$ such that $\hat\pi^{\mu\nu}(\rho_0) = 0$.

\begin{figure}[t]
\centering
\includegraphics[width=0.95\linewidth]{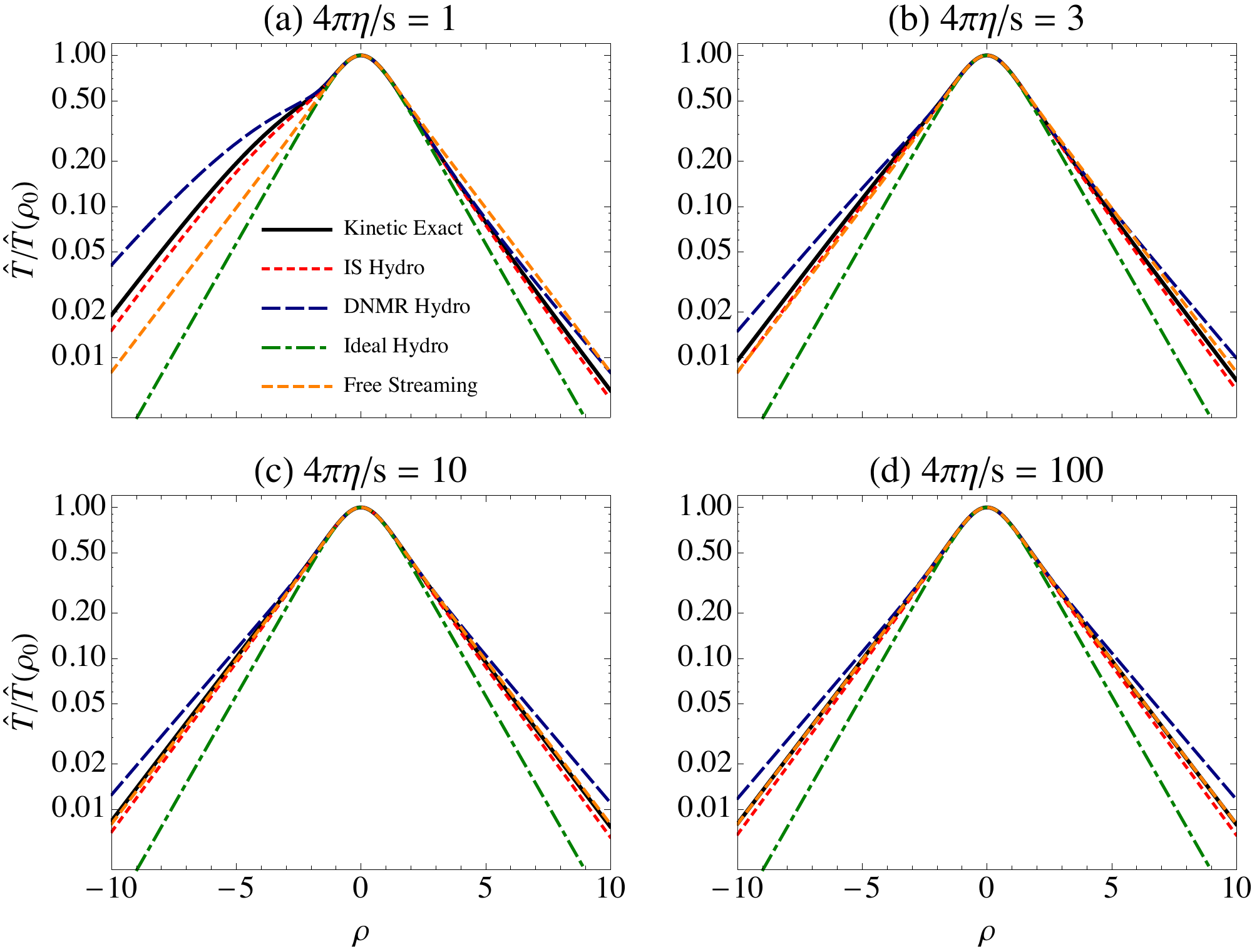}
\vspace{-3mm}
\caption{(Color online) Comparison of the de Sitter space temperature
profile obtained from the exact kinetic solution, ideal hydrodynamics, and
two second-order formulations of viscous hydrodynamics. The four panels
(a)-(d) show the results obtained assuming $4\pi\eta/\mathcal{S} = $ 1, 3, 10, and 100, respectively.
In all panels we fixed $\rho_0 = 0$ and $\hat{\cal E}(\rho_0) = 1$.}
\label{F2}
\end{figure}

\begin{figure}[t]
\centering
\includegraphics[width=0.95\linewidth]{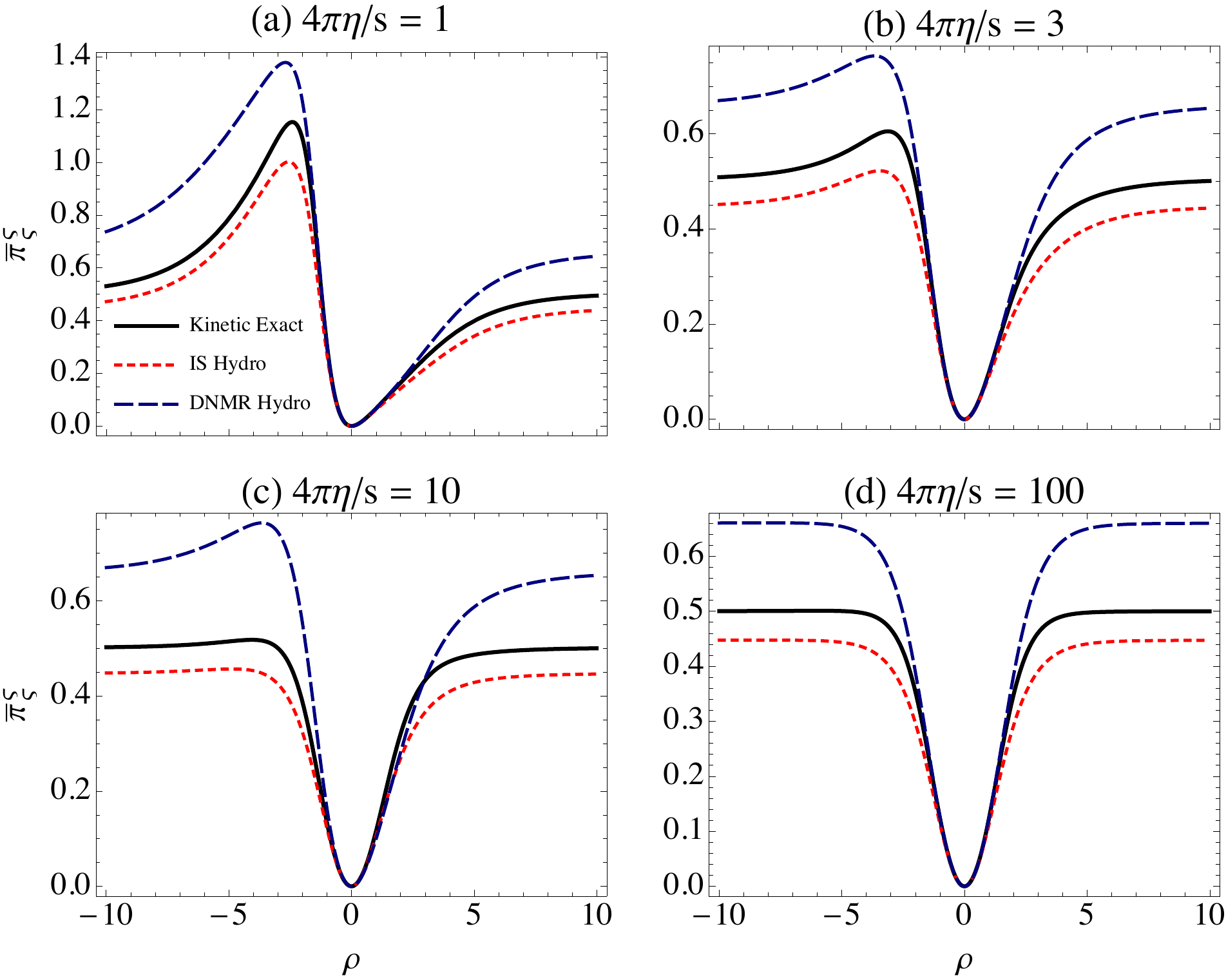}
\vspace{-3mm}
\caption{(Color online) Comparison of the normalized de Sitter 
space shear profile $\bar{\pi}_{\varsigma}^{\varsigma} \equiv%
\hat{\pi}_{\varsigma}^{\varsigma}/(\hat{T}\hat{\mathcal{S}})$ obtained from 
the exact kinetic solution and two second-order formulations of viscous  hydrodynamics. The four panels (a)-(d) show the results obtained assuming 
$4\pi\eta/\mathcal{S}{\,=\,}$1, 3, 10, and 100, respectively.
In all panels we fixed $\rho_0 = 0$ and $\hat{\cal E}(\rho_0) = 1$.}
\label{F3}
\end{figure}

\subsection{Solution in de Sitter coordinates}
\label{subsec:deSitterSol} 

In Fig.~\ref{F2} we compare the de Sitter space temperature profile 
$\hat{T}(\rho)$ for the choice $\rho_0 = 0$ with 
$\hat{\varepsilon}(\rho_0){\,=\,}1$ (corresponding to 
$\hat{T}_0=(\pi^2/3)^{1/4}{\,=\,}1.3468$) and 
$\hat\pi^{\varsigma\varsigma}(\rho_0) = 0$. The four
panels (a)-(d) show the results obtained for specific shear viscosities
$4 \pi \eta/\mathcal{S} = $1, 3, 10, and 100, respectively. In each panel we 
show the exact kinetic result as a solid black line, the IS approximation as a red short-dashed line, the DNMR
approximation as a blue long-dashed line, the ideal hydro 
approximation as a green dot-dashed line, and the free streaming 
approximation as an orange medium-dashed line. 
There are two salient points to make immediately regarding these
figures: As $\eta/\mathcal{S}$ is decreased one sees convergence to 
the ideal hydro result for positive $\rho$; as $\eta/\mathcal{S}$ is increased one observes convergence to the free streaming result for both positive and negative $\rho$; however, using the boundary condition $\hat\pi^{\mu\nu}(0) = 0$ one 
is not able to smoothly connect to the ideal limit for negative $\rho$. This 
occurs in the exact kinetic solution and in both of the second-order 
viscous hydrodynamic solutions. One can fix this problem by fine-tuning the value of $\pi^{\mu\nu}(0)$, or instead by imposing the equilibrium 
boundary condition at $\rho = -\infty$. We address this issue in more detail in
App.~\ref{app:constraints} where we show how to smoothly connect to 
the ideal hydrodynamic limit. In that same Appendix we 
comment additionally on constraints that must be satisfied by the
boundary conditions in order to obtain physically meaningful
solutions to the kinetic equation.

Focusing next on the comparison of the second-order viscous hydrodynamic
approximations to the exact kinetic solution for the de Sitter space
temperature profile in Fig.~\ref{F2}, one sees that for values of $\rho$ near 
$\rho=0$, the DNMR solution agrees better with the exact kinetic
solution; however, at large values of $\rho$ (both positive and negative) we
find the IS result to be the better approximation to the exact result.
Finally, we note that only the exact kinetic solution is able to properly
describe the largest $\eta/\mathcal{S}$ case (Fig.~\ref{F2}d), which
for all intents and purposes is the free streaming case.

In Fig.~\ref{F3} we compare the different approximations for the 
de Sitter space profile of the shear stress
$\bar\pi^{\varsigma}_\varsigma(\rho)$, again for the choice 
$\rho_0{\,=\,}0$ with $\hat{\varepsilon}(\rho_0){\,=\,}1$ and 
$\bar{\pi}^\varsigma_\varsigma(\rho_0){\,=\,}0$. The four panels (a)-(d) show 
results for four different choices of the specific shear viscosity:
$4\pi\eta/\mathcal{S}{\,=\,}$1, 3, 10, and 100. In each panel we
show the exact kinetic result as a solid black line, the IS 
hydrodynamic approximation as a short-dashed red line, and the 
DNMR approximation as a long-dashed blue line. As 
$\eta/\mathcal{S}$ is decreased, the hydrodynamic approximations 
appear to approach the exact kinetic solution; however, once 
again, although the DNMR solution seems to agree better with
the exact kinetic solution at small $\rho$, it appears to do a poorer
job than IS at large $\rho$. We note, however, that one sees quite
reasonable overall agreement between the exact kinetic solution and
the two second-order viscous hydrodynamic solutions even at extremely large 
values of $\eta/\mathcal{S}$.

\begin{figure}[b]
\includegraphics[width=0.7\linewidth]{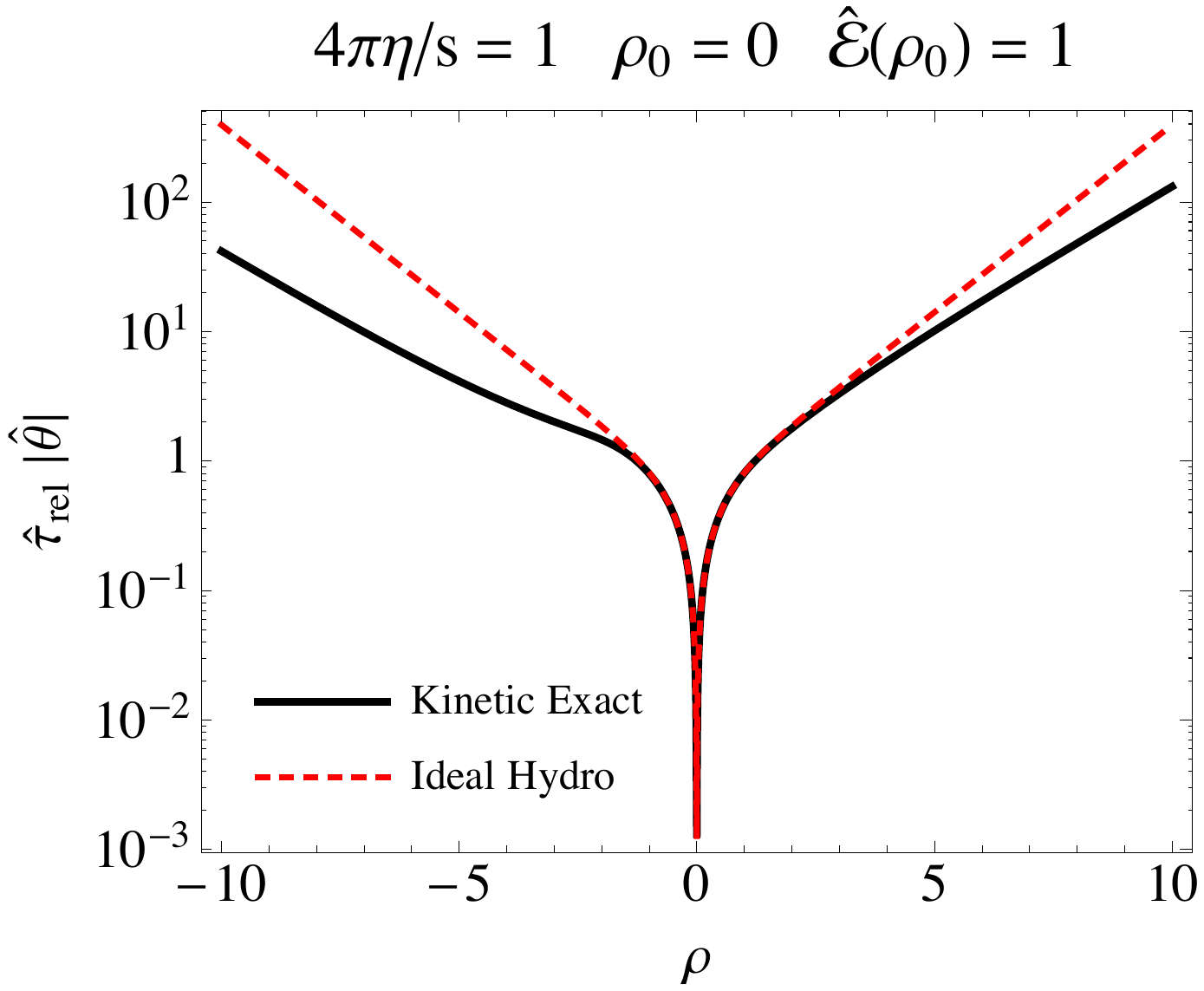}
\caption{(Color online) de Sitter time evolution of the Knudsen number
$\mathrm{Kn}$ defined in Eq.~(\ref{eq:Kn}), for the exact solution of 
the RTA Boltzmann equation (solid black line) and its ideal hydrodynamic
approximation (red dashed line). The constant $c$ in 
$\hat\tau_\mathrm{rel}{\,=\,}c/\hat{T}$ was chosen to correspond to 
$4\pi\eta/\mathcal{S}{\,=\,}1$. See text for discussion.}   
\label{F3a}
\end{figure}
%

This approximate agreement, which holds at a qualitative (${\cal O}(30\%)$) 
level even at large $|\rho{-}\rho_0|$, where for the exact solution of the RTA Boltzmann 
equation $\bar\pi^{\varsigma}_\varsigma(\rho)$ appears to approach a universal 
value of 0.5, is surprising. As we show in Fig.~\ref{F3a}, systems
with Gubser flow expand so rapidly that that they are driven away from local equilibrium
at an exponentially increasing rate as the system evolves away from the starting time
$\rho_0$ where local equilibrium initial conditions were imposed.\footnote{%
      Note that this holds for both positive and negative values of $\rho{-}\rho_0$.}
We define the Knudsen number in de Sitter coordinates as 
\begin{equation}
\label{eq:Kn}
   \mathrm{Kn} = \hat{\tau}_\mathrm{micro}/\hat{\tau}_\mathrm{macro}
    = \hat{\tau}_\mathrm{rel} \,|\hat{\theta}| 
    \equiv \hat{\tau}_\mathrm{rel} \,|\hat{\nabla}{\cdot}\hat{u}|,
\end{equation}
where $\hat\theta{\,=\,}\hat\nabla\cdot{u}$ (with $\hat\nabla_\mu$ denoting the 
covariant derivative in de Sitter coordinates) is the scalar macroscopic expansion 
rate of the Gubser flow, and $\hat\tau_\mathrm{rel}=c/\hat{T}$ is the microscopic 
relaxation time, evaluated for a $\rho$-dependent temperature $\hat{T}(\rho)$ 
that is obtained from the matching condition (\ref{eq:matchcond}) for the exact 
solution of the RTA Boltzmann equation (solid black line) and from Eq.~(\ref{eq:52}) 
for the ideal fluid approximation (dashed red line).\footnote{%
     The definition of $\mathrm{Kn}$ is independent of the coordinate system chosen, 
     and we evaluated it in de Sitter coordinates according to the last expression in 
     (\ref{eq:Kn}).}
For systems to approach local thermal equilibrium and hydrodynamics to become 
a valid approximation, $\mathrm{Kn}$ has to go to zero. Figure~\ref{F3a} 
shows that, as $|\rho{-}\rho_0|$ increases, $\mathrm{Kn}$ instead
grows exponentially, driving the system farther and farther away from local
equilibrium. The behavior of the red dashed line in Fig.~\ref{F3a} is easy
to understand: For the Gubser flow, the expansion rate in de Sitter coordinates
is
\begin{equation}
\label{eq:theta}
   \hat{\nabla}{\cdot}\hat{u} = 2\tanh\rho.
\end{equation}
Combined with $\hat{\tau}_\mathrm{rel}{\,=\,}c/\hat{T}$ and Eq.~(\ref{eq:52}), 
this yields $\mathrm{Kn}(\rho){\,=\,}(2c/\hat{T}_0)\bigl|\tanh^{1/3}(\rho)
\sinh^{2/3}(\rho)\bigr|$ which grows as $e^\rho$ for large $\rho$.\footnote{%
      These expressions assume $\rho_0{\,=\,}0$ but can obviously be 
      generalized to $\rho_0{\,\ne\,}0$.}
For the exact solution of the Boltzmann equation, viscous heating increases the
temperature relative to the ideal fluid case, tempering somewhat the rate
at which $\mathrm{Kn}$ grows for large $\rho{-}\rho_0$ without, however,
changing its exponential asymptotic behavior.

Figures~\ref{F3} demonstrates that, in spite of the exponential asymptotic growth
of the Knudsen number, the viscous stress of the system remains finite,
$\bar\pi^{\varsigma}_\varsigma(\rho)$ never growing really big. In fact, for large 
$T\tau_\mathrm{rel}$ or $\eta/\mathcal{S}$, $\bar\pi^{\varsigma}_\varsigma(\rho)$ 
never visibly exceeds its asymptotic value 0.5. This is a consequence of higher
order hydrodynamic corrections: While the growing expansion and decreasing 
scattering rates try to drive the system farther and farther away from 
local momentum isotropy, the growth of the microscopic relaxation rate 
simultaneously slows down the evolution of the viscous shear stress which 
eventually saturates at a finite value instead of following the growth of the 
dissipative force that drives it.

\subsection{Solution in Minkowski space}
\label{subsec:milneSol} 

\begin{figure}[t]
\includegraphics[width=0.46\linewidth]{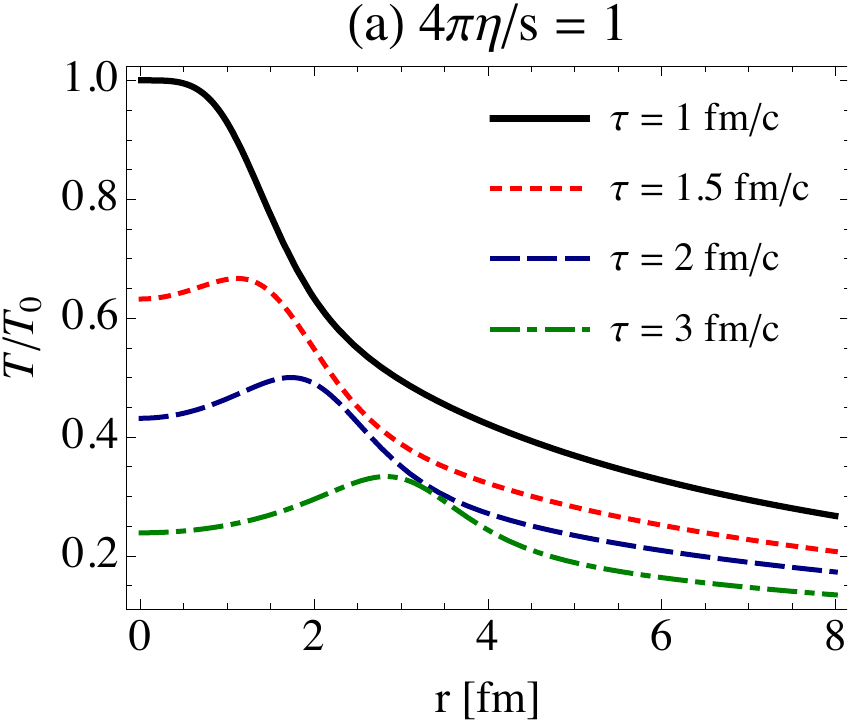}
~~~
\includegraphics[width=0.46\linewidth]{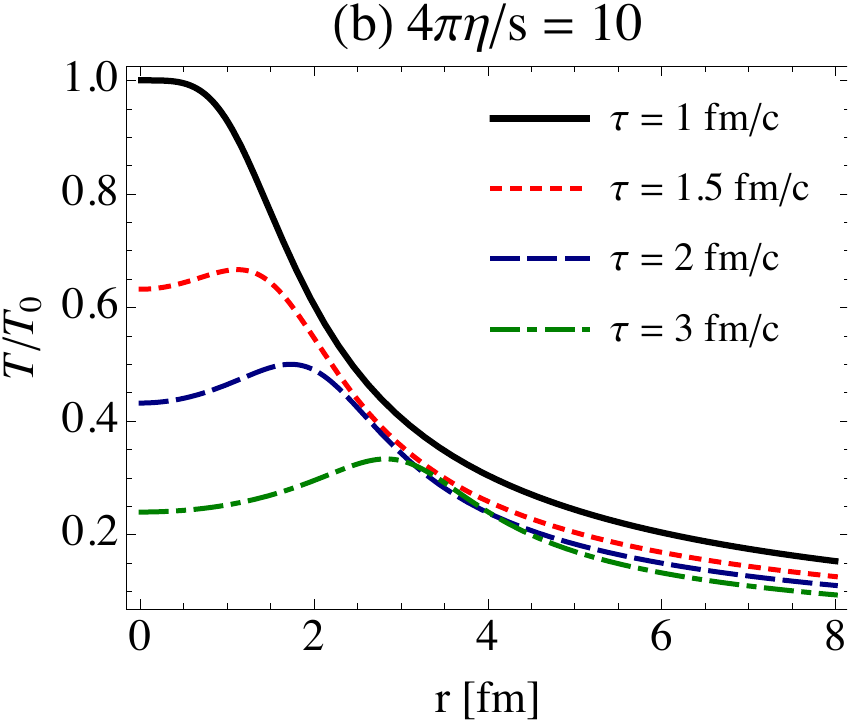}
\caption{(Color online) Exact solution for the proper-time evolution of the
temperature profile as a function of the radial coordinate $r$,
for $4 \protect\pi \protect\eta/\mathcal{S} = 1$ in panel (a) and 
$4 \protect\pi \protect\eta/\mathcal{S} = 10$ in panel (b)}.
\label{F4}
\end{figure}

Once the solution in de Sitter space for $\hat{T}$ is obtained, one can
use Eqs.~(\ref{eq:rhotheta}) and (\ref{eq:That}) to construct the solution
in Minkowski space, mapped by $r$ and $\tau$ since the system is azimuthally
symmetric and longitudinally boost invariant. For the purposes of this paper
we present results for the case $q{\,=\,}1$\,fm$^{-1}$ which corresponds
to a fairly small source size,with the ``initial'' temperature 
$\hat{T}_0\equiv\hat{T}(\rho_0{=}0)=1.3468$ (from 
$\hat{\varepsilon}_0{\,=\,}1$) translating into a temperature scale at the origin 
$r{\,=\,0}$ at a reference time $\tau_0{\,=\,}1$\,fm/$c$ of 
$T_0\equiv T(\tau_0{=}1\,\mathrm{fm}/c,r{=}0){\,\simeq\,}266$\,MeV. 
However, all plots remain unchanged under a change of the scale $q$ if we 
substitute $r\,[\mathrm{fm}]\to qr$ and $\tau\,[\mathrm{fm}/c]\to q\tau$. 
The de Sitter space results shown in Figs.~\ref{F2} and 
\ref{F3} can thus be used for any $q$ and therefore describe an entire family 
of exact solutions to the RTA Boltzmann equation and their associated 
hydrodynamic expansions with varying source size. The choice of $q$ affects the range of $\rho$ to be explored in
order to cover a given region in $r$ which increases for larger $q$ values. 
Viscous corrections, and differences between the exact microscopic and the 
approximate macroscopic evolutions will be bigger at large values of 
$\rho{-}\rho_0$.

In Fig.~\ref{F4} we show snapshots of the radial temperature profile
at four different proper times, for $4\pi\eta/S = 1$ in panel (a)
and $4\pi\eta/S = 10$ in panel (b). One sees that changing the 
shear viscosity by an order of magnitude does not seem to have 
a strong effect on the evolution of the matter near the center
(for $r\lesssim 3$ fm/$c$). However, at larger radii one notices an 
appreciable difference: for larger shear viscosity the temperature decreases more rapidly at large $r$. We note, however, that the weak dependence on the assumed value of $\eta/\mathcal{S}$ partly stems
from the fact that the flow velocity profile is here constrained by the Gubser
symmetry to be always the same, irrespective of the value of $\eta/\mathcal{S}$.

\begin{figure}[t]
\begin{centering}
\includegraphics[width=0.45\textwidth]{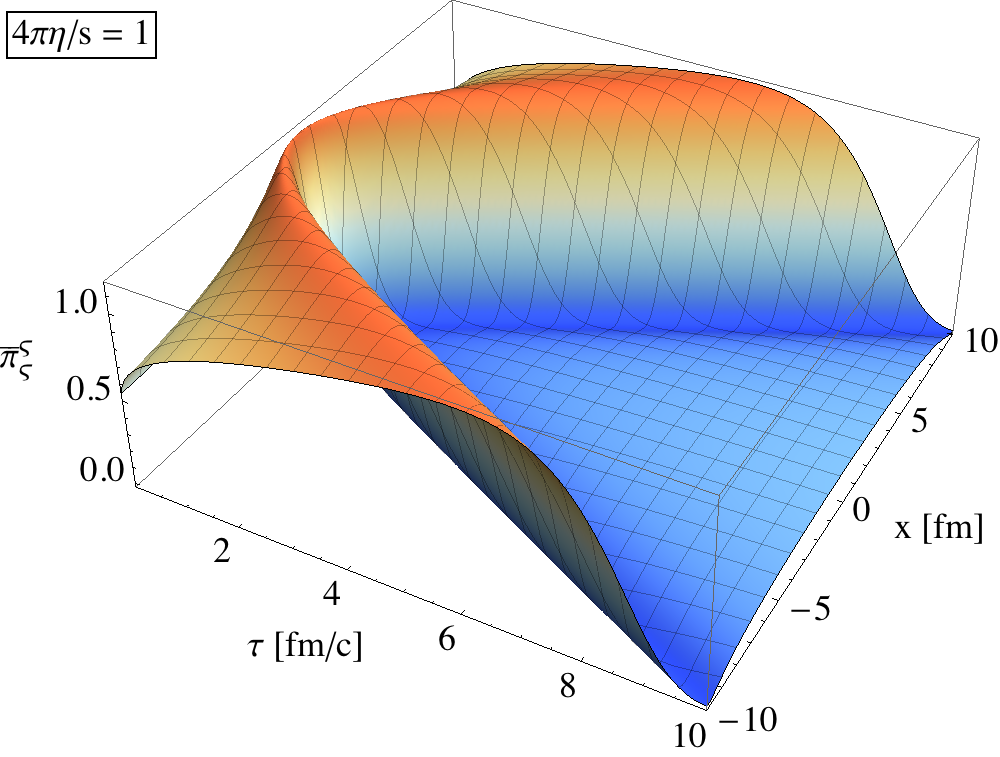}
$\;\;\;$
\includegraphics[width=0.45\textwidth]{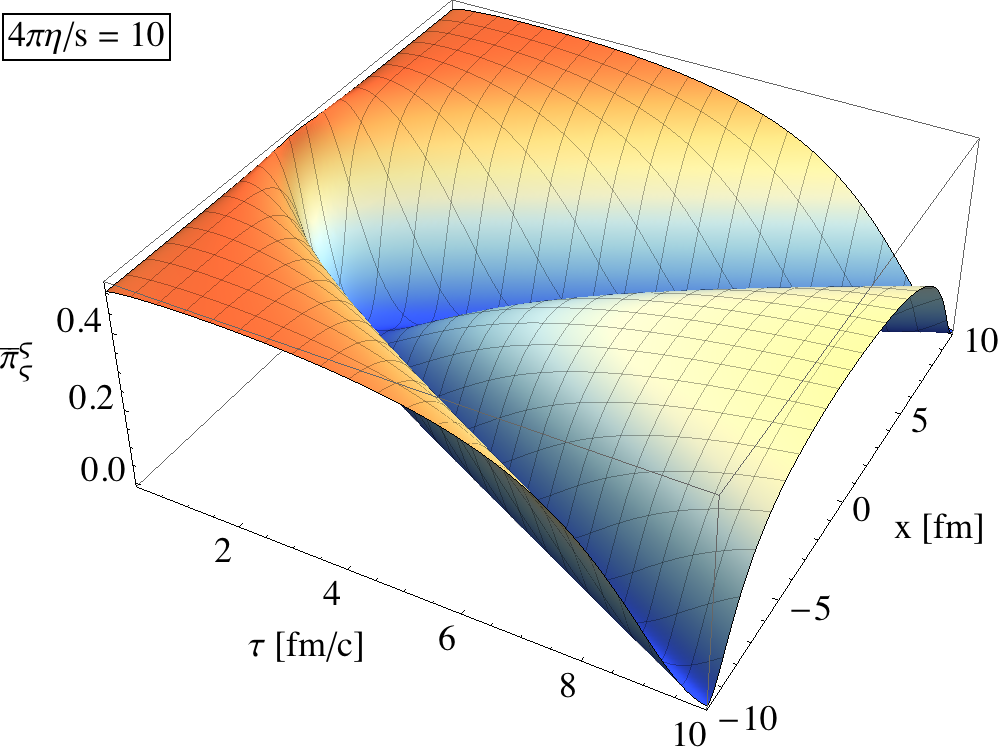}
\end{centering}
\caption{(Color online) Two-dimensional slice of the spatial and proper-time
evolution of the unitless shear stress $\bar{\pi}_{\varsigma}^{\varsigma} \equiv%
\hat{\pi}_{\varsigma}^{\varsigma}/(\hat{T}\hat{\mathcal{S}})$, 
for $4\pi\protect\eta/S = 1$ (left) and $4\pi\eta/S = 10$ (right).} 
\label{F5}
\end{figure}

The Minkowski space evolution of the scaled shear stress $\bar{\pi}_{\varsigma}^{\varsigma} \equiv%
\hat{\pi}_{\varsigma}^{\varsigma}/(\hat{T}\hat{\mathcal{S}})$ is 
plotted as a function of $x$ and $\tau$ in Fig.~\ref{F5}, for the exact solution 
of the Boltzmann equation with two different values of the specific shear viscosity. Note that the vertical scale changes between the left and right figures. Also note, that although we explicitly show the $x$-dependence the solution is cylindrically symmetric by construction.
As this figure shows, the assumed value of $\eta/\mathcal{S}$ has a strong 
effect on the spacetime evolution of the shear stress.

\begin{figure}[t]
\centering
\includegraphics[width=0.98\textwidth]{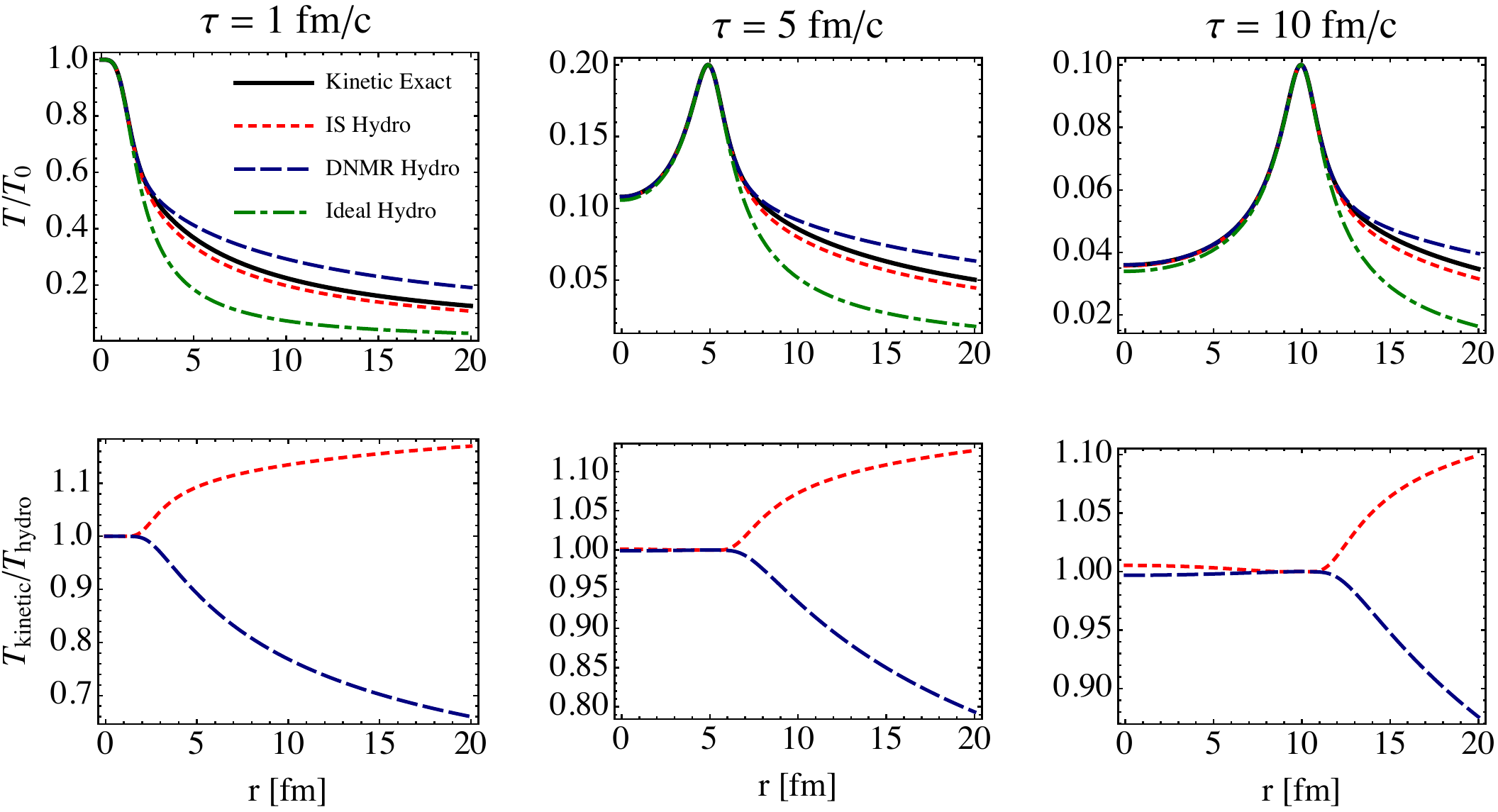}
\caption{(Color online) Snapshots of the temperature profile in Milne 
coordinates obtained from the exact kinetic solution (solid black line), 
ideal
hydrodynamics (dot-dashed green line), the second-order IS solution (red
short-dashed line), and the second-order DNMR solution (long-dashed blue
line). For this figure we assumed $4 \protect\pi \protect\eta/\mathcal{S} = 1
$.}
\label{F6}
\end{figure}

\begin{figure}[t]
\centering
\includegraphics[width=0.98\textwidth]{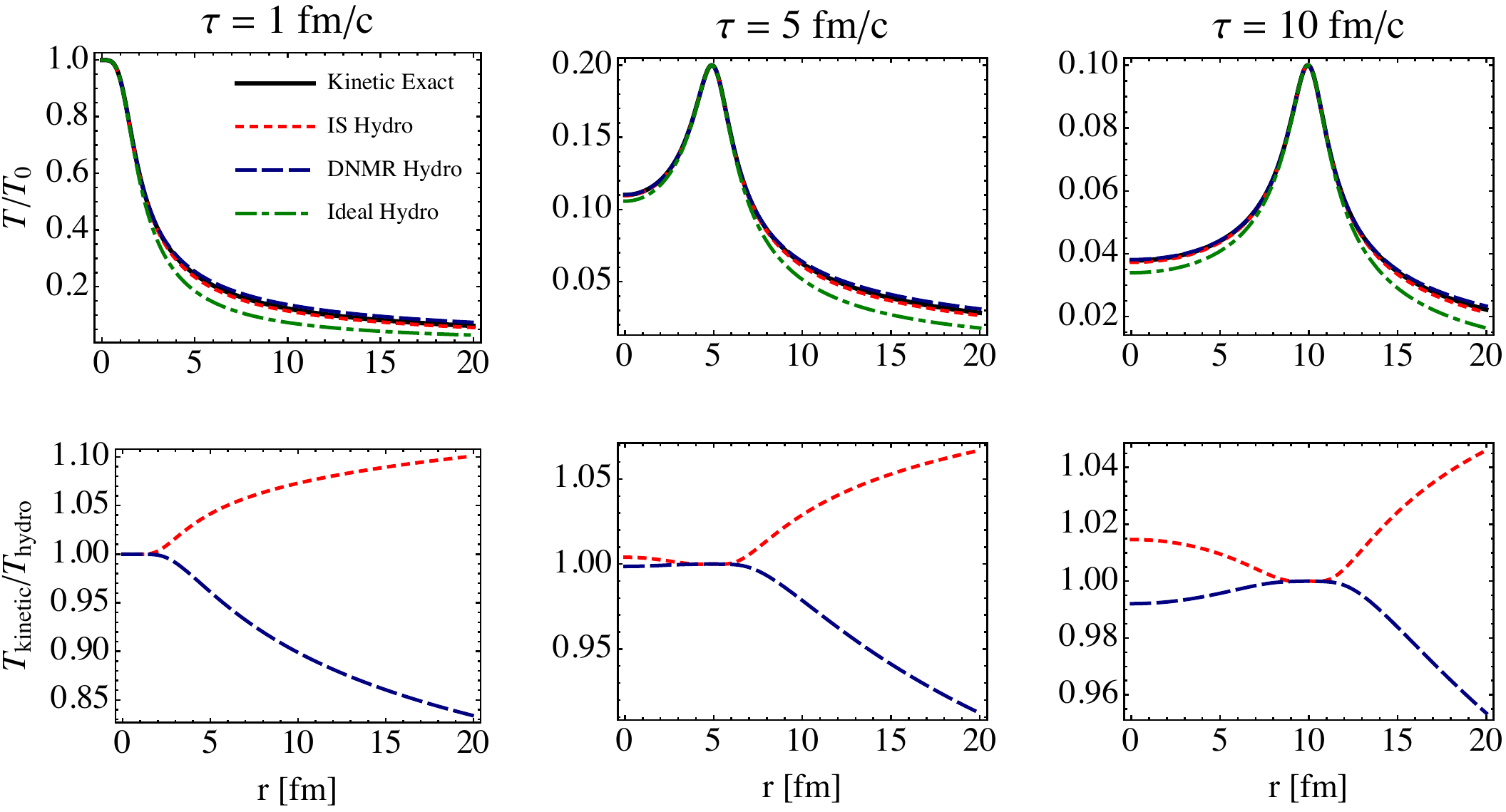}
\caption{(Color online) Same as Fig.~\ref{F6}, but for 
$4\protect\pi \protect\eta/\mathcal{S} =10$.}
\label{F7}
\end{figure}

Finally, we compare in Fig.~\ref{F6} (for 
$\eta/\mathcal{S}{\,=\,}1/(4\pi)$) and Fig.~\ref{F7} (for 
$\eta/\mathcal{S}{\,=\,}10/(4\pi)$) snapshots of the 
Minkowski space temperature profile obtained from the exact kinetic solution 
(solid black line) with different hydrodynamic approximations: ideal 
hydrodynamics (dot-dashed green line), second-order IS viscous 
hydrodynamics (red short-dashed line), and the DNMR 
approximation to second-order viscous hydrodynamics (long-dashed blue 
line). In the top set of panels we show radial profiles at three different 
longitudinal proper times, $\tau{\,=\,}1,\ 5,$ and 10\,fm/$c$.
In the bottom set of panels we show the ratio between the 
temperatures corresponding to the exact kinetic result and those of 
the two second-order viscous hydrodynamic approximations. Note 
that even at the reference time $\tau_0{\,=\,}1$\,fm/$c$ the temperature 
profiles are not the same: since we impose initial conditions not at a 
fixed longitudinal proper time in Minkowski space, but at a fixed ``de Sitter 
time'' $\rho_0$, there is nothing special about the time $\tau{\,=\,}1$\,fm/$c$ 
(except that it is the natural longitudinal proper time scale for a scale 
parameter $q{\,=\,}1$\,fm$^{-1}$). Figs.~\ref{F6} and \ref{F7} show that
the temperatures corresponding to the exact and approximate 
solutions always agree at their peak value, and that the position of this  
peak (which corresponds to the initial value $\rho_0{\,=\,}0$ in de Sitter space)
moves out in the radial direction along the line 
$r{\,=\,}\sqrt{\tau^2{-}1}$ as $\tau$ increases (see Eq.~\eqref{definerho}). 

From Fig.~\ref{F6} we see that for $4\pi\eta/\mathcal{S}{\,=\,}1$
the maximum error in the temperature from the IS approach
is on the order of 10-15\% in the $(\tau,r)$ region shown in the plots,
and somewhat larger for the DNMR approximation. Somewhat counterintuitively,  
Fig.~\ref{F7} seems to show a significantly smaller error for a 10 times larger
shear viscosity. However, more careful inspection reveals that this is only the case to the
right of the peak, which corresponds to negative $\rho$ values; in the central
region ($r\lesssim\sqrt{\tau^2{-}1}$) which corresponds to positive $\rho$
values, the late-time differences between the exact kinetic solution and the 
second-order viscous hydrodynamic approaches {\em increase} with 
increasing $\eta/\mathcal{S}$, with the DNMR approach giving slightly better 
agreement with the exact kinetic solution. In general, the deviations of the 
hydrodynamic approximations from the exact solution appear to be larger
at negative than for positive $\rho$ values; this agrees qualitatively with the
pattern observed in Figs.~\ref{F2} and \ref{F3}. Not surprisingly, the ideal 
hydrodynamic approximation fares worst in all cases.

\section{Conclusions}
\label{sec:concl} 

In this paper we presented an exact solution of the Boltzmann equation in
the relaxation time approximation for a system that expands with 
Gubser symmetric longitudinal and transverse flow. We showed that, in the 
conformal (massless) limit, the Boltzmann equation has an emergent Weyl 
symmetry. Transforming to de Sitter coordinates and imposing the Gubser 
flow as the four-velocity profile as well as other constraints imposed
by the Gubser symmetry on the allowed dependences of the distribution 
function on the phase-space coordinates, we were able to cast the Boltzmann equation into stationary form. This allowed us to 
solve it in the form of one-dimensional integral equations
for the full distribution function, temperature profile, and all components 
of the energy-momentum tensor. From these integral equations we
could analytically extract the ideal hydrodynamic solution, several
variants of second-order viscous hydrodynamic solutions, and the 
free-streaming solution.

The resulting one-dimensional integral equations were then
solved numerically using an iterative method which allowed us to obtain an
exact solution to the kinetic equation to arbitrary numerical accuracy in de
Sitter space. The resulting exact de Sitter space solution can be
analytically mapped back to Minkowski space and can be used to describe an
entire family of exact solutions in this space with varying physical source
size. For a given source size corresponding to the choice $q{\,=\,}1$\,fm$^{-1}$,
we then made quantitative comparisons between the different hydrodynamic
approximations and the exact kinetic solution. We found that, while not 
perfect, the second-order hydrodynamic approximations gave reasonable 
results even in the limit of large specific shear viscosities 
$\eta /\mathcal{S}$.

One complication in these comparisons is that, in order to 
preserve the Gubser symmetry, initial conditions must be implemented at constant ``de Sitter time'' $\rho$ in de Sitter space. When mapped back to 
Minkowski space, one cannot guarantee that the exact and approximate solutions have the same radial temperature profile at a fixed
longitudinal proper time. While this introduces some subtleties into the interpretation of the comparisons, it does not detract 
from the fact that one is now able to construct exact solutions to the 
Boltzmann equation for systems that feature simultaneous (albeit 
still highly symmetric) longitudinal and transverse expansion, irrespective 
of the assumed value of $\eta/\mathcal{S}$ (or, equivalently, the relaxation time $\tau_\mathrm{rel}$). Looking forward, it will be interesting to compare the solutions described in this work with
higher-order truncations of viscous hydrodynamics and with anisotropic hydrodynamics. Moreover, using similar techniques it 
should be possible to find additional exact solutions to the Boltzmann 
equation for other relativistically expanding systems (featuring e.g. transversally 
anisotropic (2+1)-dimensional flow) by considering more general conformal 
maps between Minkowski space and other curved spacetimes \cite{Hatta:2014gqa,Hatta:2014gga}. 
We leave this for future work.

\acknowledgments{
M.M. thanks G. Chirilli, J.~McEwen, and Z.L.~Carson for useful discussions on topics related to differential geometry and conformal field theory. G.S. Denicol was supported by a Banting Fellowship from the Natural Sciences and Engineering Research Council of Canada.  U.H. and M.M. were supported by the U.S. Department of Energy, Office of Science, Office of Nuclear Physics under Award No.~\rm{DE-SC0004286}. J.N.\ thanks the Conselho Nacional de Desenvolvimento Cient\'ifico e Tecnol\'ogico (CNPq) and Funda\c c\~ao de Amparo \`a Pesquisa do Estado de S\~ao Paulo (FAPESP) for support. U.H. and M.S. were supported in part (in the framework of the JET Collaboration) by U.S. DOE Awards No.~\rm{DE-SC0004104} and \rm{DE-AC0205CH11231}. M.S. would also like to thank the Institute for Theoretical Physics, Johann Wolfgang Goethe-Universit\"at, Frankfurt, and the Institute for Theoretical Physics, Technische Universit\"at Wien, for hosting him during the final stages of this project. Finally, U.H., M.M. and J.N. acknowledge support through a bilateral scientific exchange program between the Office of Sponsored Research at The Ohio State University and FAPESP. 
}

\appendix

\section{Conformal hydrodynamics}
\label{app:IS} 

In this Appendix we derive the fluid-dynamical equations of motion in de
Sitter coordinates with the metric $\hat g_{\mu \nu }=\mathrm{diag}(-1,\cosh^2{\rho},\cosh^2\rho\sin^2\theta,1)$. In these coordinates our system is static,
{\it i.e.} $\hat u_{\mu }=\left(-1,0,0,0\right) $, and the equations of motion simplify
considerably and can be solved with little numerical effort. The nonzero
components of the Christoffel symbol in these coordinates are 
$\hat\Gamma _{\theta \theta }^{\rho}=\cosh\rho\sinh\rho$, 
$\hat\Gamma _{\phi\phi }^\rho=\cosh\rho\sinh\rho\sin^2\theta$, 
$\hat\Gamma _{\rho\theta}^\theta=\hat\Gamma _{\theta\rho}^\theta
=\hat\Gamma _{\rho\phi}^\phi=\tanh\rho$, 
$\hat\Gamma _{\phi\phi}^\theta=-\sin\theta\cos\theta$, 
and $\hat\Gamma_{\theta\phi}^\phi=\hat\Gamma _{\phi\theta}^\phi=\cot\theta$. 
Also, the determinant of the metric is $\sqrt{-\hat g}=\cosh^2\rho\sin\theta$.

The expansion rate $\hat\theta =\hat{\nabla}_{\mu }\hat u^{\mu }$ is therefore given by
\begin{equation}
\hat\theta =\frac{1}{\sqrt{{-}\hat g}}\hat\partial _{\mu }\left( \sqrt{- \hat g}\hat u^{\mu }\right)
=2\tanh \rho  
\end{equation}
($\hat\nabla_\mu$ is the general relativistic covariant derivative in de Sitter coordinates), while the shear tensor $\hat\sigma _{\mu \nu }=\hat\Delta _{\mu\nu}^{\alpha\beta} \hat\nabla_\alpha\hat u_\beta$ can be shown to be
\begin{equation}
\hat\sigma _{\mu \nu} =\hat\Gamma _{\mu \nu }^{\rho }-\frac{1}{3}\hat\Delta _{\mu\nu}
\hat\theta =\mathrm{diag}\left(0,\frac{1}{3}\cosh\rho\sinh\rho,
\frac{1}{3}\sin^2\theta\cosh\rho\sinh\rho, -\frac{2}{3}\tanh\rho \right).
\end{equation}
The projection operators $\hat\Delta _{\mu \nu }$ and 
$\hat\Delta _{\mu\nu }^{\alpha\beta }$ were defined in 
the main text of the paper.

The energy conservation equation can then be re-expressed as
\begin{equation}
\hat u^{\mu }\hat\nabla_{\mu }\hat\varepsilon 
+\bigl(\hat\varepsilon{+}\hat{\mathcal{P}}\bigr) \hat\nabla_{\mu }\hat u^{\mu}
+\hat\pi^{\mu \nu }\hat\sigma _{\mu \nu }=0
\Longrightarrow \partial _{\rho}\hat\varepsilon 
+ 2\bigl(\hat\varepsilon{+}\hat{\mathcal{P}}\bigr)\tanh\rho 
- \hat\pi^\varsigma_\varsigma \tanh\rho =0,
\end{equation}
where we used the tracelessness $\hat\pi _{\mu }^{\mu }=0$. Since for a conformal fluid 
$\hat\varepsilon\sim\hat T^{4}$, once can rewrite this equation as an equation of
motion for the temperature:
\begin{equation}
\frac{1}{\hat T}\partial _{\rho }\hat T
=-\frac{2}{3}\tanh \rho +\frac{1}{3}\frac{\hat\pi^{\varsigma\varsigma}}
                                                               {\hat\varepsilon{+}\hat{\mathcal{P}}}\tanh\rho.
\end{equation}

The equation for the shear stress tensor is written as
\begin{equation}
\hat\tau_{\pi}\hat\Delta _{\mu \nu}^{\alpha\beta}\,
\hat u^{\lambda }\hat\nabla_\lambda\hat\pi_{\alpha\beta } + \hat\pi_{\mu\nu}
=-2\hat\eta \hat\sigma _{\mu \nu }-\frac{4}{3}\hat\pi _{\mu\nu}\hat\theta 
-\frac{10}{7}\hat\pi _{\left\langle\mu\right.}^{\lambda}
\hat\sigma_{\left.\nu\right\rangle\lambda}.
\end{equation}
The term $\hat\pi_{\left\langle\mu\right.}^\lambda\hat\sigma_{\left.\nu\right\rangle\lambda}$ 
can be simplified as follows:
\begin{equation}
\hat\pi _{\left\langle\mu\right.}^\lambda\hat\sigma_{\left.\nu \right\rangle\lambda}
=\Delta_{\mu\nu}^{\alpha\beta}\hat\pi_\alpha^\lambda\hat\sigma_{\beta\lambda}
=\frac{1}{2}\hat\pi_\mu^\lambda \hat\sigma_{\nu\lambda}
+\frac{1}{2}\hat\pi_\nu^\lambda\hat\sigma_{\mu\lambda}
+\frac{1}{3}\hat\Delta_{\mu\nu}\hat\pi^{\varsigma\varsigma}\tanh\rho,  
\end{equation}
where
\begin{eqnarray}
\hat\pi _{\left\langle \varsigma \right. }^{\lambda }\hat\sigma _{\left. \varsigma
\right\rangle \lambda } &=&-\frac{1}{3}\hat\pi_\varsigma^\varsigma \tanh\rho , 
\nonumber \\
\hat\pi_{\left\langle \theta \right.}^\lambda\hat\sigma_{\left.\theta\right\rangle\lambda} 
&=&-\frac{1}{3}\cosh\rho\sinh\rho \hat\pi _\phi^\phi,  
\nonumber \\
\hat\pi_{\left\langle\phi\right.}^\lambda\hat\sigma _{\left.\phi\right\rangle\lambda} 
&=&\frac{1}{3}\cosh\rho \sinh\rho \sin^2\theta \hat\pi_\theta^\theta.
\end{eqnarray}
The relaxation term $\hat\Delta _{\mu \nu}^{\alpha\beta}\hat u^\lambda\hat\nabla_\lambda
\hat\pi_{\alpha\beta}$ is worked out as
\begin{equation}
\hat\Delta _{\mu \nu}^{\alpha\beta}\hat u^\lambda \hat\nabla_\lambda \hat\pi_{\alpha\beta}
=\hat{D}\hat\pi _{\mu\nu}-\hat\pi_{\alpha\beta}\hat{D}\hat\Delta_{\mu\nu}^{\alpha\beta},
\end{equation}
where $\hat{D}=\hat u^\lambda\hat\nabla_\lambda$. The first term on the right is given 
by
\begin{equation}
\hat{D}\hat\pi _{\mu\nu} =\hat u^\lambda \hat\nabla_\lambda \hat\pi_{\mu \nu}
=\hat u^\lambda \hat\partial_\lambda \hat\pi_{\mu\nu} 
-\hat u^\lambda\hat\Gamma_{\mu\lambda}^\alpha \hat\pi_{\alpha\nu}
-\hat u^\lambda\hat\Gamma_{\nu\lambda}^\alpha \hat\pi_{\alpha\mu}
\end{equation}
while the second one is 
\begin{equation}
-\hat\pi_{\alpha\beta}\hat{D}\hat\Delta_{\mu\nu}^{\alpha\beta}
=-\hat\pi_{\alpha\beta} \hat{D}\bigl(\hat\Delta_\mu^\alpha\hat\Delta_\nu^\beta\bigr) 
=\left( \hat u_{\nu}\hat\pi _{\mu }^{\alpha }{+}\hat u_{\mu }\hat\pi _{\nu}^{\alpha }\right) 
\hat u^{\sigma }\hat\Gamma _{\sigma \alpha }^{\lambda }\hat u_{\lambda}\,. 
\end{equation}
Thus, the equations of motion for $\hat\pi_{\theta\theta}$, $\hat\pi_{\phi\phi}$, 
and $\hat\pi_{\varsigma\varsigma}$ become
\begin{eqnarray}
\hat\tau _{\pi }\partial _{\rho }\hat\pi _{\theta \theta }+\hat\pi _{\theta \theta } 
&=&-\left(\frac{2}{3}\hat\eta \cosh^2\rho 
       +\frac{2}{3}\hat\tau _{\pi }\hat\pi _{\theta\theta}
       +\frac{10}{21}\frac{\hat\tau _{\pi }\hat\pi _{\phi \phi }}{\sin^2\theta}\right)\tanh\rho, 
\\
\hat\tau _{\pi }\partial_{\rho }\hat\pi _{\phi \phi }+\hat\pi _{\phi \phi } 
&=&-\left(\frac{2}{3}\hat\eta \cosh^2\rho\sin^2\theta 
       +\frac{2}{3}\hat\tau _{\pi }\hat\pi _{\phi \phi }
       +\frac{10}{21}\hat\tau _{\pi }\hat\pi_{\theta \theta }\sin^2\theta\right)\tanh\rho, \quad
\\\label{eq:a13}
\hat\tau _{\pi }\partial _{\rho }\hat\pi _{\varsigma \varsigma }+\hat\pi _{\varsigma \varsigma} 
&=& \left(\frac{4}{3}\hat\eta -\frac{8}{3}\hat\tau _{\pi }\hat\pi _{\varsigma \varsigma}
              +\frac{10}{21}\hat\pi_{\varsigma \varsigma}\right)\tanh\rho .
\end{eqnarray}
One sees that $\hat\pi _{\varsigma \varsigma}$ decouples while the remaining two
components of the shear stress are coupled to each other. Due to the tracelessness of
the shear stress tensor it is, however, sufficient to only solve the last equation, Eq.~(\ref{eq:a13}). 

Defining the scaled shear stress $\bar{\pi}_\varsigma^\varsigma=\hat\pi_\varsigma^\varsigma/(\hat\varepsilon{+}\hat{\mathcal{P}})$
and using the fact that in massless kinetic theory 
$\hat\eta/\hat\tau_\pi=(\hat\varepsilon{+}\hat{\mathcal{P}})/5$, we arrive after a few more steps at the following final form of the equations of motion in de Sitter coordinates:
\begin{eqnarray}
\label{eq:a14}
\frac{1}{\hat T}\partial _{\rho }\hat T+\frac{2}{3}\tanh\rho
&=&\frac{1}{3}\bar{\pi}_{\varsigma}^{\varsigma }\tanh\rho, 
\\\label{eq:a15}
\partial _{\rho }\bar{\pi}_{\varsigma}^{\varsigma}
+\frac{\bar{\pi}_{\varsigma }^{\varsigma}}{\hat\tau _{\pi }}\tanh \rho 
+\frac{4}{3}\left( \bar{\pi}_{\varsigma}^{\varsigma}\right) ^{2} 
&=&\frac{4}{15}\tanh \rho +\frac{10}{7}\bar{\pi}_{\varsigma}^{\varsigma}\tanh\rho.
\end{eqnarray}
In traditional Israel-Stewart theory \cite{Marrochio:2013wla}, where the term proportional to 
$\hat\pi_{\left\langle \mu \right. }^{\lambda }\hat\sigma _{\left. \nu \right\rangle\lambda }$ is absent, the equation of motion for $\hat\pi^{\varsigma\varsigma}$ reduces to
\begin{equation}
\label{eq:a16}
\partial _{\rho }\bar{\pi}_{\varsigma}^{\varsigma}+\frac{\bar{\pi}_{\varsigma}^{\varsigma
}}{\hat\tau _{\pi }}\tanh \rho +\frac{4}{3}\left( \bar{\pi}_{\varsigma}^{\varsigma
}\right) ^{2}=\frac{4}{15}\tanh \rho .
\end{equation}
The last three equations are Eqs.~(\ref{eq:istemp})-(\ref{eq:dnmrshear}) in Sec.~\ref{sec:result}.

\section{Physical constraints on the de Sitter space boundary condition}
\label{app:constraints} 

As mentioned in the body of this paper, in order to obtain the exact
solution to the kinetic equation one must specify an initial condition in de
Sitter space. The exact kinetic solution obtained herein assumed  
$\hat\pi^{\varsigma}_\varsigma(\rho_0) = 0$ at some
particular value $\rho_0$. In the Sec.~\ref{sec:result} we chose 
$\rho_0=0$ in order to make the comparison between the various 
approaches most transparent, however, one has some degree of freedom 
in the choice of this parameter. One issue with the solutions presented in 
the main body is that it is not possible to take the limit 
$\eta/\mathcal{S} \rightarrow 0$ in order to recover the
ideal hydrodynamics limit (see Fig.~\ref{F2}). As can be seen from
Fig.~\ref{F2}, for negative $\rho$ the solution does not converge to
the ideal hydrodynamic limit as $\eta/\mathcal{S} \rightarrow 0$. In fact,
for very small values of $\eta/\mathcal{S}$ one finds that the solution
diverges at some finite negative $\rho$. This behavior is not restricted to
the exact kinetic solution and occurs within both the IS and DNMR
second-order viscous hydrodynamic approaches as well.

In order to take the small $\eta/\mathcal{S}$ limit, one must very carefully
take $\eta/\mathcal{S} \rightarrow 0$ using a positive value of 
$\hat\pi^{\varsigma}_\varsigma(0)$ which vanishes only when $\eta/\mathcal{S}$ is precisely zero. For the IS and DNMR second-order 
viscous hydrodynamic solutions, one can iteratively determine the necessary 
value of $\hat\pi^{\varsigma}_\varsigma(\rho_0)$ required to recover 
the ideal hydrodynamic result as $\eta/\mathcal{S} \rightarrow 0$; however, 
due to the form of the initial distribution function assumed
herein, it is not currently possible to implement a finite value for 
$\hat\pi^{\varsigma}_\varsigma(\rho_0)$ in the exact kinetic solution. As an
alternative approach which guarantees convergence to the ideal hydrodynamic
result as $\eta/\mathcal{S} \rightarrow 0$ one could instead fix the
boundary condition on the left edge of the simulation region.%
\footnote{\label{fn11}%
   As the simulation region is enlarged, this corresponds to fixing the
   boundary condition at $\rho \to -\infty$.}
If this is done, one can straightforwardly take the ideal limit.

\begin{figure}[t]
\includegraphics[width=0.43\textwidth]{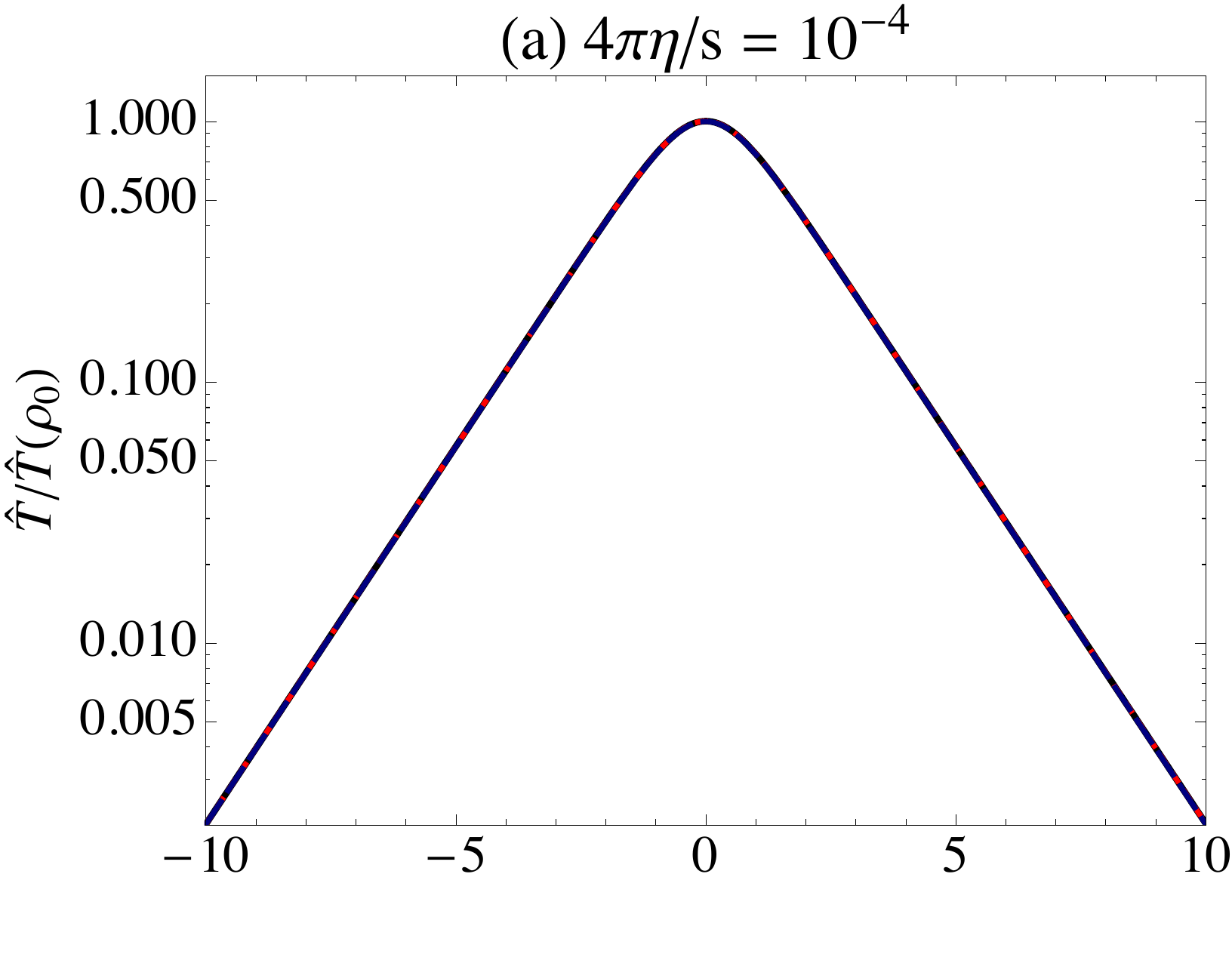}
\includegraphics[width=0.43\textwidth]{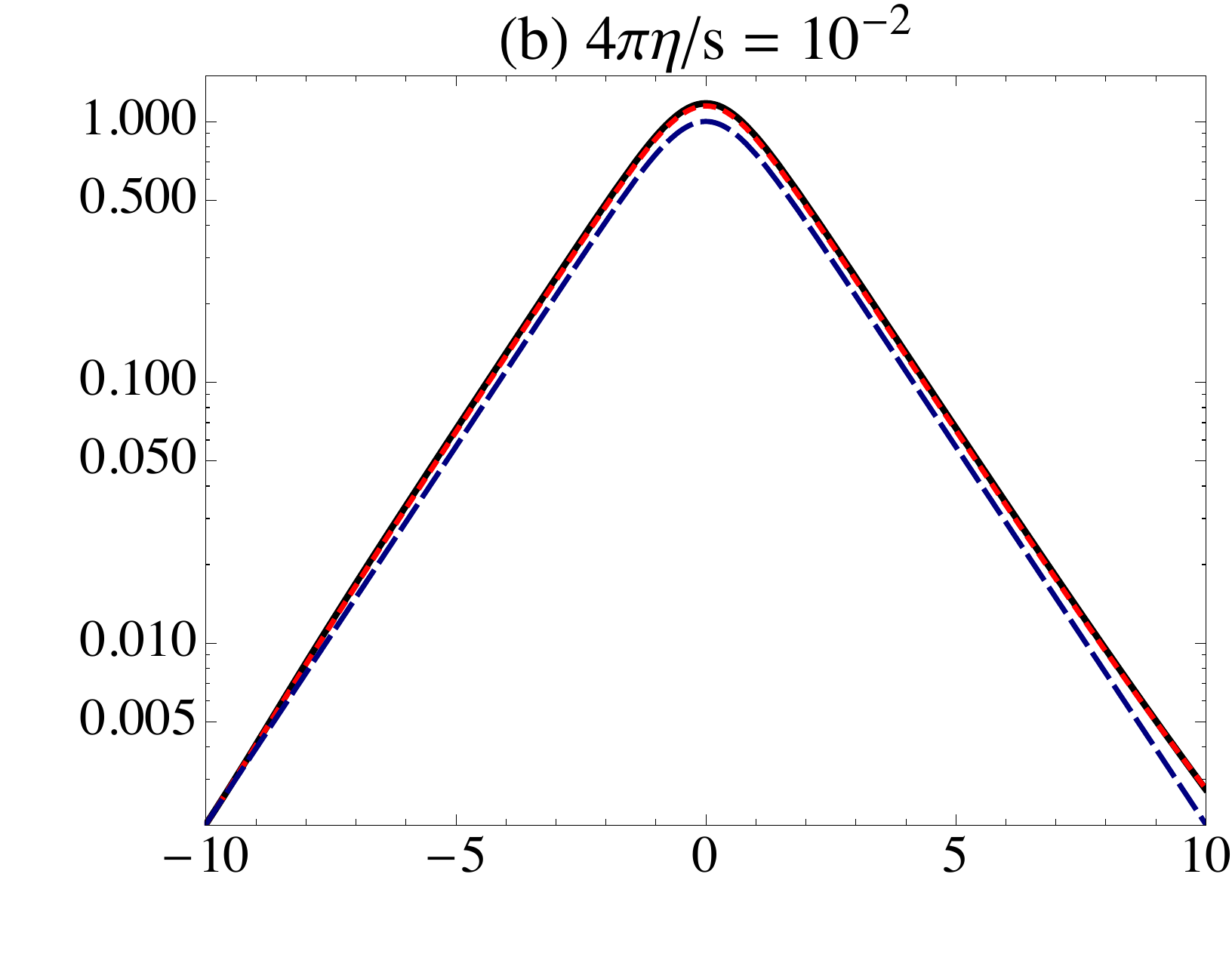}
\newline

\vspace{-10mm}
\includegraphics[width=0.43\textwidth]{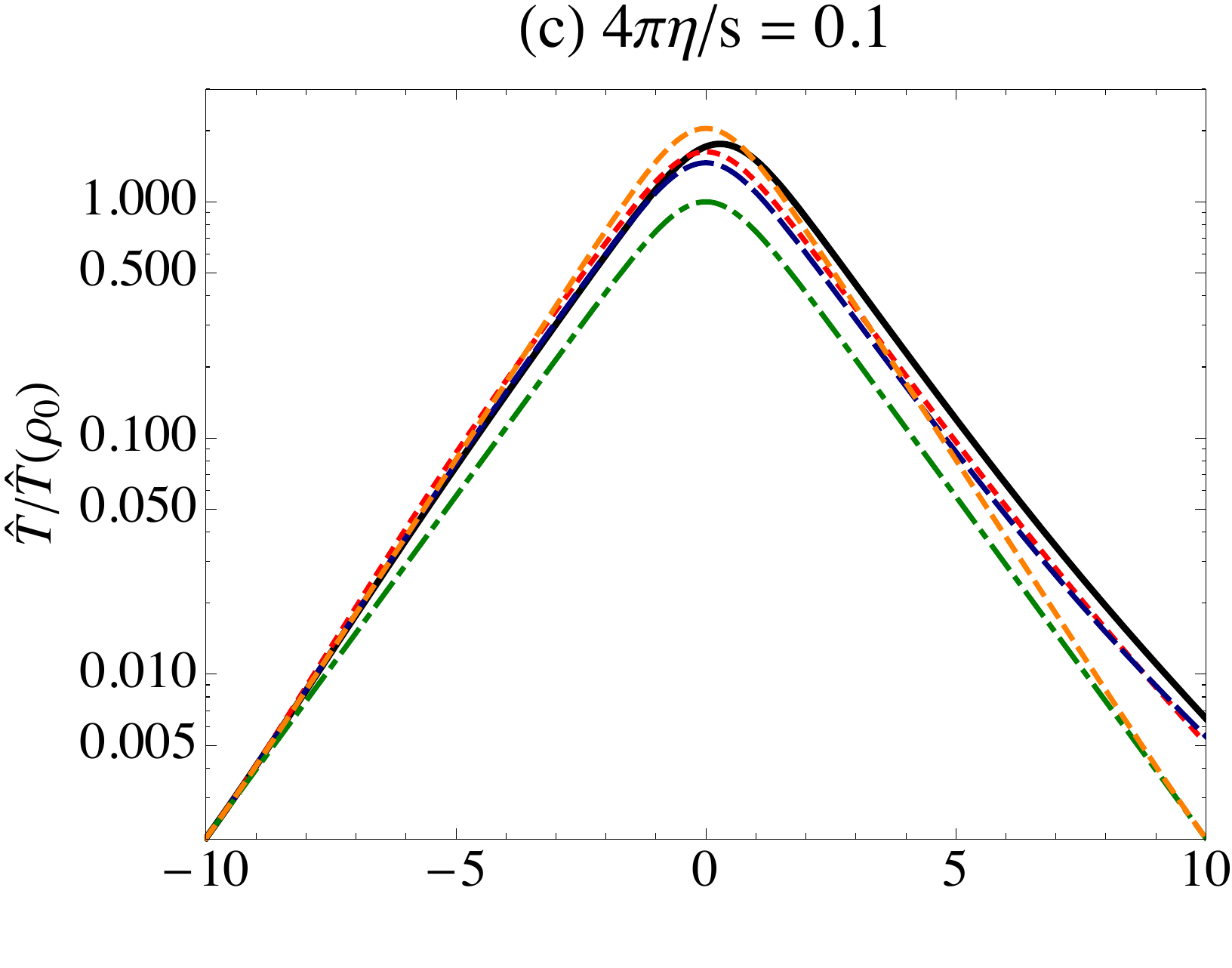}
\includegraphics[width=0.43\textwidth]{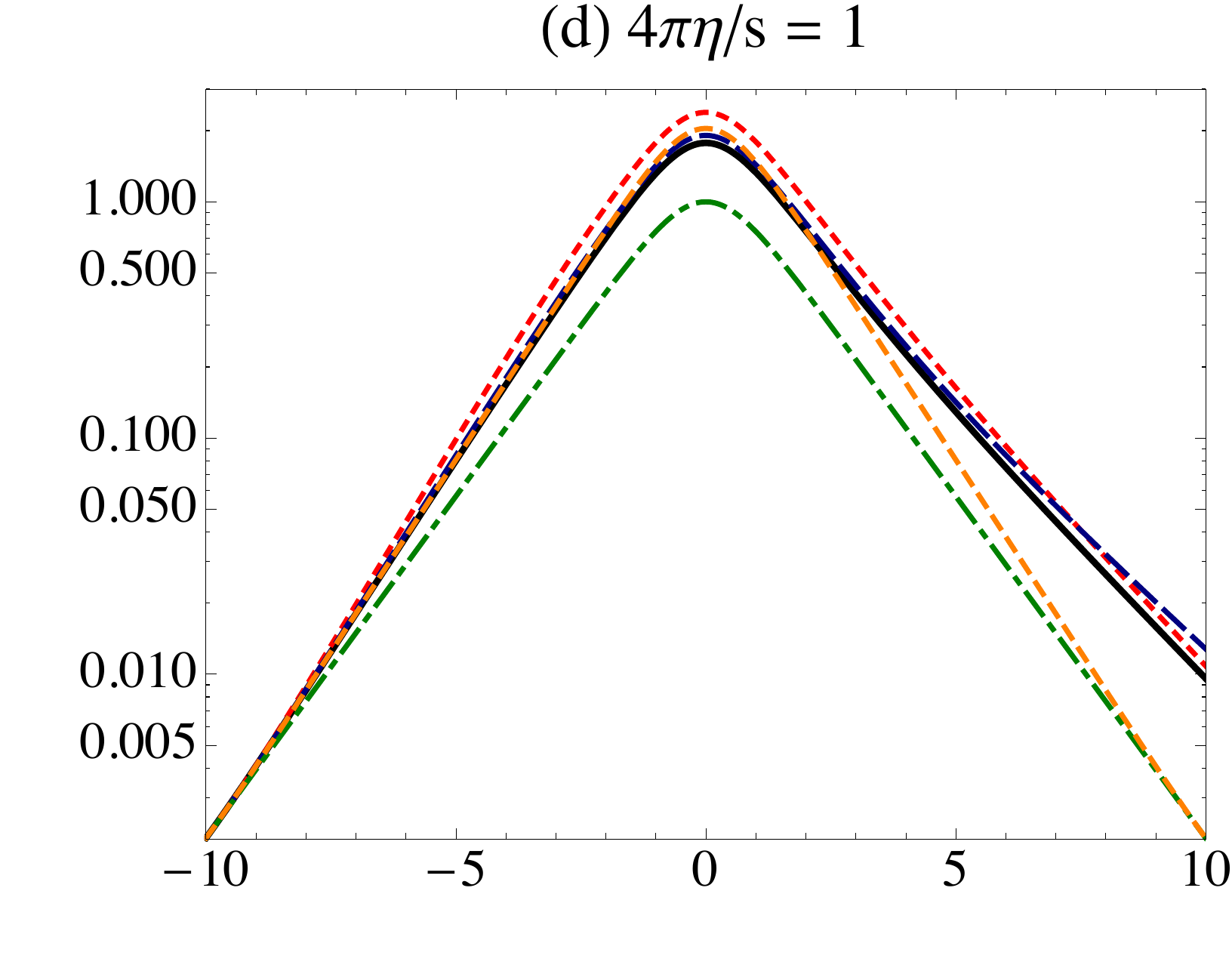}
\newline

\vspace{-10mm}
\hspace*{-5mm}\includegraphics[width=0.43\textwidth]{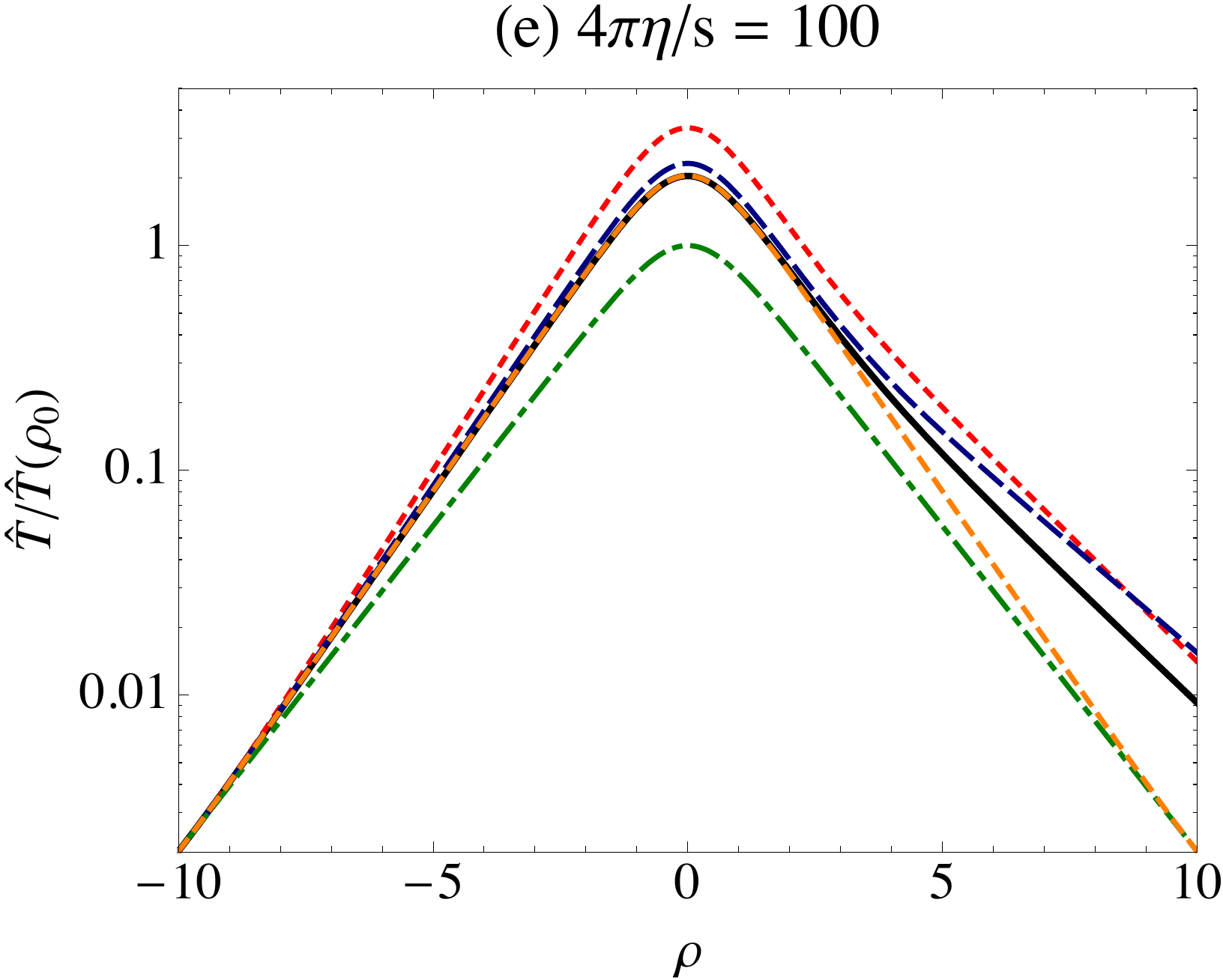}
\includegraphics[width=0.43\textwidth]{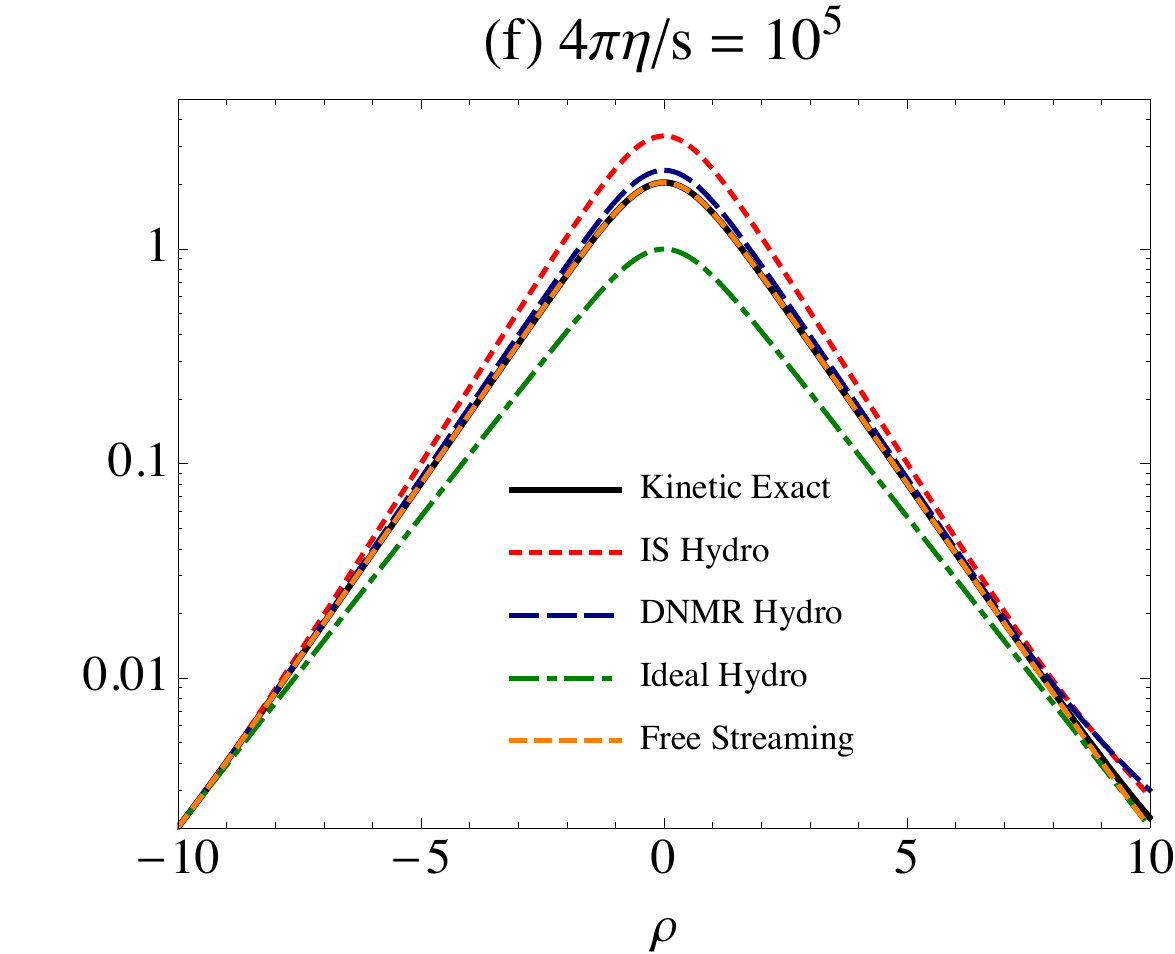}
\vspace{-4mm}
\caption{(Color online) Comparison of the de Sitter space temperature
profile obtained from the exact kinetic solution, ideal hydrodynamics, and
two second-order formulations of viscous hydrodynamics, with
equilibrium initial conditions imposed at $\rho_0{\,=\,}{-}10$ with $\hat{T}(\rho_0) = 2.02018\times10^{-3}$. Panels (a)-(f) show the results obtained assuming 
$4\pi\eta/\mathcal{S}{\,=\,}$10$^{-4}$, 10$^{-2}$, 1, 10, 100, and $10^5$, respectively.
\label{F8}}
\end{figure}

\begin{figure}[t]
\includegraphics[width=0.47\textwidth]{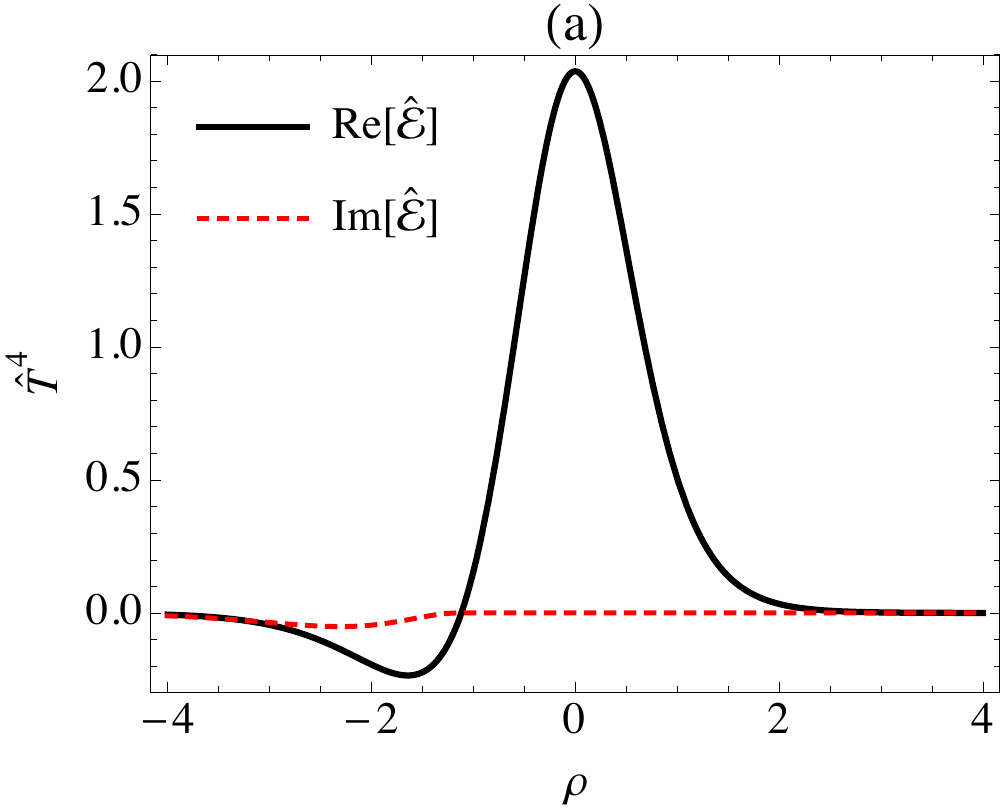}
~~
\includegraphics[width=0.47\textwidth]{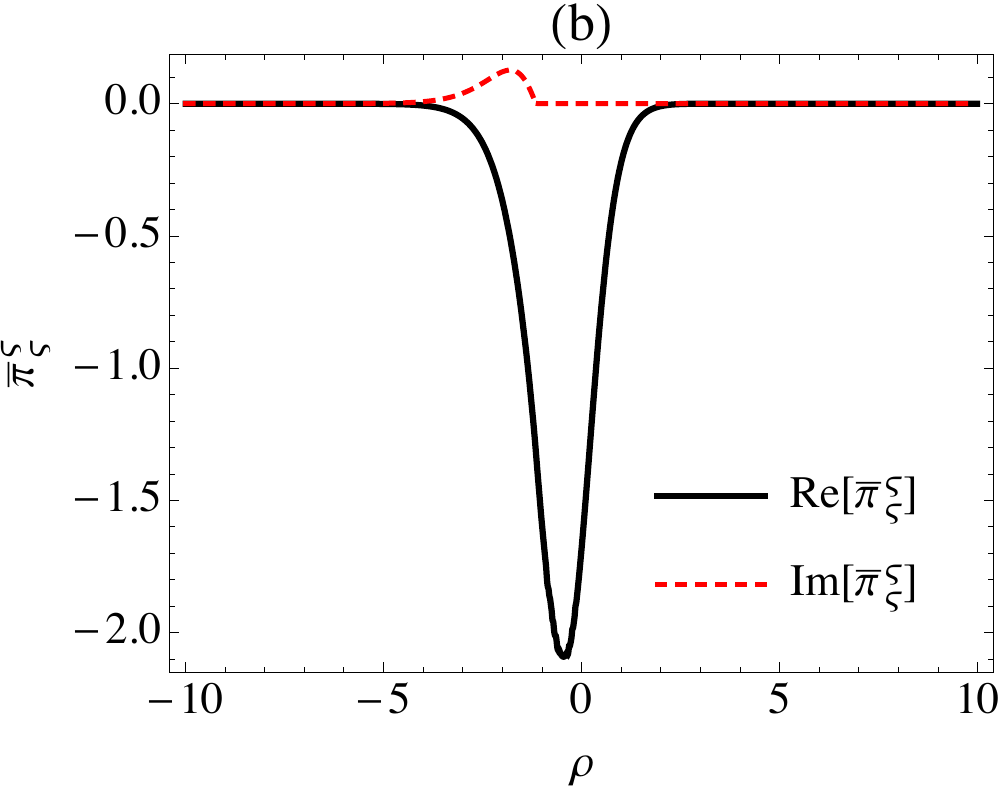}
\caption{(Color online) The de Sitter space temperature profile obtained
from the exact kinetic solution assuming $4 \protect\pi \protect\eta/%
\mathcal{S} = $ 3 and $\protect\rho_0 = 3$ with $\hat{T}(\protect\rho_0) = 0.214477$.}
\label{F9}
\end{figure}

To demonstrate this, in Fig.~\ref{F8} we plot the results
obtained with such a boundary condition for both small and large values of 
$\eta/\mathcal{S}$. For very small specific shear viscosities the 
exact kinetic solution is computationally very demanding and cannot
be obtained with our computing resources, so  in panels (a) and (b) we 
graph only our macroscopic solutions for ideal and viscous hydrodynamics. 
Panel (a) shows that, with equilibrium boundary conditions implemented
at large negative de Sitter times, the two viscous hydrodynamic approximations 
(which in Fig.~\ref{F2} are seen to bracket the exact kinetic solution at
negative values of $\rho{-}\rho_0$) perfectly reproduce the ideal fluid limit 
when $\eta/\mathcal{S}$ becomes very small. Fig.~\ref{F9}f, on the other 
hand, demonstrates that for very large values of $\eta/\mathcal{S}$ 
the exact kinetic solution converges perfectly to the free-streaming limit. 
For finite shear viscosity, however, a complication with this type of 
boundary condition is that by implementing it at the left edge in 
de Sitter space one will not find the same temperature and shear correction 
at positive $\rho$ values (which in Minkowski space map to the central 
fireball region at times on the order of 1 fm/$c$). As a result, it is more 
difficult to make apples-to-apples comparisons of the Minkowski-space 
evolution in this case.
It is for this reason that in the body of the text we chose the simpler 
``initial condition'' $\hat\pi^{\varsigma}_\varsigma{\,=\,}0$ 
at $\rho_0{\,=\,}0$. This condition guarantees that all 
approaches start with an isotropic initial condition that is essentially 
free from shear corrections near the fireball center at longitudinal 
proper times on the order of 1\,fm/$c$.

We note also that, for any value of $\eta/\mathcal{S}$ one could also
attempt to initialize the system in de Sitter space in equilibrium at
a positive value of $\rho_0$; however, one finds in practice that doing this 
can result in complex-valued energy densities and, if $\rho_0$ is taken to 
be large enough, the numerical solution will fail to converge. Once again, this
behavior is not unique to the kinetic solution and similar behavior can be
seen in the second-order viscous hydrodynamic solutions. Of course,
negative or complex energy densities are unphysical, and this numerical 
phenomenon indicates that the physical range of $\rho$ values, where the
distribution function corresponding to a thermal equilibrium boundary
condition at $\rho_0$ remains positive definite for all momenta, ends 
somewhere at sufficiently large negative values of $\rho{-}\rho_0$. We 
have observed this phenomenon even for $\rho_0{\,=\,}0$ where it appears 
to happen at larger and larger negative values of $\rho$ as 
$\eta/\mathcal{S}$ decreases, but never completely disappears. 

As a concrete illustration of this, in Fig.~\ref{F9} we plot the de Sitter space 
profile of $\hat{T}^4$ in panel (a) and $\bar\pi^\varsigma_\varsigma$
in panel (b). In both panels, the black solid line is the real part of the quantity 
and the red short-dashed line is the imaginary part of the quantity. In the case 
shown, the code can be made to converge to arbitrary accuracy, however, the
resulting solutions are complex-valued for sufficiently negative 
$\rho{-}\rho_0$. While this result is, in fact, a mathematical solution 
to the RTA Boltzmann equation subject to the Gubser flow profile, 
a complex temperature is physically meaningless and (we believe) 
indicative of deeper underlying problems related to a violation of positivity of 
the distribution function $f$ at large negative $\rho{-}\rho_0$. As a minimum,
one must therefore restrict the choice of $\rho_0$ such that (a) the code converges and (b) both the temperature and shear correction remain 
real-valued over the entire de Sitter space domain considered.%
\footnote{\label{fn12}%
   We caution that, since the solution for the temperature (which enters all
   other computed quantities) is obtained from a moment of the distribution
   function (see Eq.~\eqref{eq:efftemp}), problems with the positivity of the
   distribution function in some parts of momentum space may not 
   immediately signal themselves through complex temperature values. 
   Therefore, unphysical behaviour of the distribution function may remain 
   hidden in part of the $\rho$ space determined by this procedure.} 
In future work we plan to relax the requirement 
$\hat{\pi}^\varsigma_\varsigma(\rho_0){\,=\,}0$ in which case it
may be possible to impose initial conditions over a larger range of $\rho_0$,
without the solutions becoming unphysical.

\bibliography{confBoltz}

\end{document}